\documentclass[a4paper,twocolumn,accepted=2023-09-19]{quantumarticle}
\pdfoutput=1
\usepackage[latin9]{inputenc}
\usepackage{epsfig}
\usepackage{verbatim}
\usepackage{ucs}
\usepackage{bbm}
\usepackage{amsmath}
\usepackage{amssymb}
\usepackage{hyperref}

\newcommand{\bra}[1]{\langle #1|}
\newcommand{\ket}[1]{|#1\rangle}
\newcommand{\ketbra}[1]{| #1\rangle \langle #1|}

\newcommand{\be}{\begin{equation}}
\newcommand{\ee}{\end{equation}}
\newcommand{\eea}{\end{eqnarray}}
\newcommand{\bea}{\begin{eqnarray}}

\newcommand{\va}[1]{\ensuremath{(\Delta#1)^2}}
\newcommand{\vasq}[1]{\ensuremath{[\Delta#1]^2}}

\newcommand{\ex}[1]{\ensuremath{\langle{#1}\rangle}}
\newcommand{\exs}[1]{\ensuremath{\langle{#1}\rangle}}

\newcommand{\qed}{\ensuremath{\hfill \blacksquare}}

\newcommand{\kommentar}[1]{}
\newcommand{\trace}{{\rm Tr}}

\newcommand{\forget}[1]{}

\newcommand{\EQ}[1]{Eq.~\eqref{#1}}
\newcommand{\EQS}[1]{Eqs.~\eqref{#1}}
\newcommand{\EQL}[1]{Equation~\eqref{#1}}
\newcommand{\SEC}[1]{Sec.~\ref{#1}}
\newcommand{\FIG}[1]{Fig.~\ref{#1}}
\newcommand{\REF}[1]{Ref.~\cite{#1}}
\newcommand{\REFS}[1]{Refs.~\cite{#1}}

\newcommand{\APP}[1]{Appendix~\ref{#1}}

\newcommand{\OBS}[1]{Observation~\ref{#1}}
\newcommand{\OBSS}[1]{Observations~\ref{#1}}
\newcommand{\DEFOBS}[1]{{\bf Observation \refstepcounter{observation}\theobservation\label{#1}.} }
\newcounter{observation}

\newcommand{\PROOF}{{\it Proof.} }

\newcommand{\EXAMPLE}[1]{Example~\ref{#1}}
\newcommand{\DEFEXAMPLE}[1]{{\bf Example \refstepcounter{example}\theexample\label{#1}.} }
\newcounter{example}

\newcommand{\DEFINITION}[1]{Definition~\ref{#1}}

\newcommand{\DEFDEFINITION}[1]{{\bf Definition \refstepcounter{definition}\thedefinition\label{#1}.} }
\newcounter{definition}

\newcommand{\Hamiltonian}{\ensuremath{H}}

\begin{document}

\title{Quantum Wasserstein distance based on an optimization over separable states}

\author{G\'eza T\'oth}
\orcid{0000-0002-9602-751X}
\email{toth@alumni.nd.edu}
\affiliation{Theoretical Physics, University of the Basque Country UPV/EHU, ES-48080 Bilbao, Spain}
\affiliation{EHU Quantum Center, University of the Basque Country UPV/EHU, Barrio Sarriena s/n, ES-48940 Leioa, Biscay, Spain}
\affiliation{Donostia International Physics Center (DIPC),  ES-20080 San Sebasti\'an, Spain}
\affiliation{IKERBASQUE, Basque Foundation for Science, ES-48011 Bilbao, Spain}
\affiliation{Institute for Solid State Physics and Optics, Wigner Research Centre for Physics, HU-1525 Budapest, Hungary}

\author{J\'ozsef Pitrik}
\email{pitrik@math.bme.hu}
\affiliation{Institute for Solid State Physics and Optics, Wigner Research Centre for Physics, HU-1525 Budapest, Hungary}
\affiliation{Alfr\'ed R\'enyi Institute of Mathematics, Re\'altanoda u. 13-15., HU-1053 Budapest, Hungary}
\affiliation{Department of Analysis and Operations Research, Institute of Mathematics, Budapest University of Technology and Economics, M\H{u}egyetem rkp. 3., HU-1111 Budapest, Hungary}

\begin{abstract}
We define the quantum Wasserstein distance such that the optimization of the coupling is carried out over bipartite separable states rather than bipartite quantum states in general, and examine its properties. Surprisingly, we find that the self-distance is related to the quantum Fisher information. We present a transport map corresponding to an optimal bipartite separable state. We discuss how the quantum Wasserstein distance introduced is connected to criteria detecting quantum entanglement. We define variance-like quantities that can be obtained from the quantum Wasserstein distance  by replacing the minimization over quantum states by a maximization. We extend our results to a family of generalized quantum Fisher information quantities.
\\
\vskip0.05cm
\noindent {\rm Dedicated to the memory of D\'enes Petz on the occasion of his 70$^{th}$ birthday.}
\end{abstract}

\maketitle

\section{Introduction}

A classical Wasserstein distance is a metric between probability distributions $\mu$ and $\nu,$ 
induced by the problem of optimal mass transportation \cite{Monge1781Memoire,Kantorovitch1958Translocation}. It reflects the minimal effort that is required in order to morph the mass of $\mu$ into the mass distribution of $\nu$. Methods based on the theory of optimal transport and advantageous properties of Wasserstein metrics have achieved great success in several important fields of pure mathematics including probability theory \cite{Boissard2015Distributions,Butkovsky2014Subgeometric}, theory of (stochastic) partial differential equations \cite{Hairer2011Asymptotic,Hairer2008Spectral}, variational problems \cite{Figalli2010Mass,Figalli2011Shape} and geometry of metric spaces \cite{LottVillani2009Ricci,RenesseSturm2005Transport,Sturm2006Geometry,Sturm2006Geometry2}. In recent years, there have been a lot of results concerning the description of isometries of Wasserstein spaces, too \cite{Kloeckner2010Geometric,Geherr2019,Viro2020,Viro2021,Viro2022,Kiss2022,Geher2023}. Moreover, optimal transport and Wasserstein metric are also used as tools in applied mathematics. In particular, there are applications to image and signal processing \cite{Kolouri2016Radon,Wang2013Linear,Kolouri2017Optimal}, medical imaging \cite{Gramfort2015Fast,Su2015Shape} and machine learning \cite{Arjovsky2017Wasserstein,Moselhy2012Bayesian,Peyre2019Computational,Frogner2015Learning,Ramdas2017Wasserstein,Srivastava2018Scalable}.

The non-commutative generalization, the so-called quantum optimal transport has been at the center of attention, as it lead to the definition of several new and very useful notions in quantum physics. The first, semi-classical approach of \.Zyczkowski and Slomczynski, has been motivated by applications in quantum chaos \cite{Zyczkowski1998TheMonge,Zyczkowski2001TheMonge,Bengtsson2006Geometry}.  The method of Biane and Voiculescu is related to free probability \cite{Biane2011Free}, while the one of Carlen, Maas, Datta, and Rouz\'e \cite{CarlenMaas2014Analog,CarlenMaas2017Gradient,CarlenMaas2020Non-commutative,DattaRouze2019Concentration,DattaRouze2020Relating} is based on a dynamical interpretation. Caglioti, Golse, Mouhot, and Paul presented an approach based on a static interpretation \cite{Golse2016On,Golse2017The,Golse2018Wave,Golse2018TheQuantum,Caglioti2020Quantum,Caglioti2021Towards}. Finally De Palma and Trevisan used quantum channels \cite{DePalma2021Quantum}, and De Palma, Marvian, Trevisan, and Lloyd defined the quantum earth mover's distance, i.e., the quantum Wasserstein distance order 1 \cite{DePalma2021TheQuantum}, while Bistron, Cole, Eckstein, Friedland and \.Zyczkowski formulated a quantum Wasserstein distance based on an antisymmetric cost function \cite{Friedland2022Quantum,Friedland2021Quantum,Bistron2022Monotonicity}.

One of the key results of quantum optimal transport is the definition of the quantum Wasserstein distance \cite{Zyczkowski1998TheMonge,Zyczkowski2001TheMonge,Bengtsson2006Geometry,Golse2016On,Golse2017The,Golse2018Wave,Golse2018TheQuantum,DePalma2021Quantum,DePalma2021TheQuantum,Caglioti2020Quantum,Caglioti2021Towards,Geher2023Quantum,Li2022Wasserstein}. It has the often desirable feature that it is not necessarily maximal for two quantum states orthogonal to each other, which is beneficial, for instance, when  performing learning on quantum data \cite{Kiani2022Learning}. Some of the properties of the new quantities are puzzling, yet point to profound relations between seemingly unrelated fields of quantum physics. For instance, the quantum Wasserstein distance order 2 of the quantum state from itself can be nonzero, while in the classical case the self-distance is always zero. In particular, as we have mentioned, the quantum Wasserstein distance has been defined based on a quantum channel formalism \cite{DePalma2021Quantum}, and it has been shown that the square of the self-distance is equal to the Wigner-Yanase skew information of the quantum state \cite{Wigner1963INFORMATION}.  At this point the questions arises: Can other fields of quantum physics help to interpret these results? For example, is it possible to relate the findings above to entanglement theory \cite{Horodecki2009Quantum,Guhne2009Entanglement,Friis2019} such that we obtain new and meaningful relations naturally?

Before presenting our results, let us summarize the definitions of the quantum Wasserstein distance.

\DEFDEFINITION{def:D}The square of the distance between two quantum states described by the density matrices $\varrho$ and $\sigma$ is given by De Palma and Trevisan as \cite{DePalma2021Quantum}
\begin{align}
D_{\rm DPT}(\varrho,\sigma)^2=\frac 1 2
\min_{ \varrho_{12} }\sum_{n=1}^N\;&
\trace[(\Hamiltonian_n^T\otimes \openone-\openone\otimes \Hamiltonian_n)^2  \varrho_{12} ],\nonumber\\
\textrm{s.~t. }&
\varrho_{12}\in\mathcal D,\nonumber\\
& {\rm Tr}_2(\varrho_{12})=\varrho^T, \nonumber\\
& {\rm Tr}_1(\varrho_{12})=\sigma,\label{eq:distance}
\end{align}
where $A^T$ denotes the matrix transpose of $A,$ and $H_1, H_2, ...,H_N$ are Hermitian operators \footnote{We use the convention that in the expression within the trace in \EQ{eq:distance}, $\Hamiltonian_n\otimes \openone$ denotes an operator in which $\Hamiltonian_n$ acts on subsystem $1$ and $\openone$ acts on subsystem $2.$}, while $\mathcal D$ is the set of density matrices, i.e., Hermitian matrices fulfilling
\begin{equation}
\varrho_{12} =\varrho_{12}^\dagger, {\rm Tr}(\varrho_{12})=1, \varrho_{12}\ge 0.
\end{equation}

In this approach, there is a bipartite density matrix $\varrho_{12},$ called coupling, corresponding to any transport map between $\varrho$ and $\sigma,$ and vice versa, there is a transport map corresponding to any coupling \cite{DePalma2021Quantum}. Moreover, it has been shown that for the self-distance of a state  \cite{DePalma2021Quantum}
\begin{equation}
D_{\rm DPT}(\varrho,\varrho)^2=\sum_{n=1}^N I_\varrho(\Hamiltonian_n)\label{eq:DrhoI}
\end{equation}
holds,  where the Wigner-Yanase skew information is defined as \cite{Wigner1963INFORMATION}
\begin{equation}
I_{\varrho}(\Hamiltonian)={\rm Tr}(\Hamiltonian^2\varrho)-{\rm Tr}(\Hamiltonian\sqrt{\varrho}\Hamiltonian\sqrt{\varrho}).
\end{equation}
This profound result connects seemingly two very different notions of quantum physics, as it has been mentioned in the introduction.

The Wasserstein distance has also been defined in a slightly different way.

\DEFDEFINITION{def:DGMCP}Golse, Mouhot, Paul and Caglioti  defined the square of the distance as \cite{Golse2016On,Caglioti2021Towards,Golse2018TheQuantum,Golse2017The,Golse2018Wave,Caglioti2020Quantum}
\begin{align}
D_{\rm GMPC}(\varrho,\sigma)^2\quad\quad\quad\nonumber\\=\frac 1 2
\min_{ \varrho_{12}} \sum_{n=1}^N\;&
\trace[(\Hamiltonian_n\otimes \openone-\openone\otimes \Hamiltonian_n)^2  \varrho_{12} ],\nonumber\\
\textrm{s.~t. }&
\varrho_{12}\in\mathcal D,\label{eq:GMPC_distance}  \nonumber\\
& {\rm Tr}_2(\varrho_{12})=\varrho, \nonumber\\
& {\rm Tr}_1(\varrho_{12})=\sigma.
\end{align}

In this paper, we will obtain new quantities by restricting the optimization to separable states in the above definitions. We will show that, in this case, the square of the self-distance equals the quantum Fisher information times a constant, while in \EQ{eq:DrhoI} it was related to the Wigner-Yanase skew information. The quantum Fisher information is a central quantity in quantum estimation theory and quantum metrology, a field that is concerned with metrological tasks in which the quantumness of the system plays an essential role  \cite{Giovannetti2004Quantum-Enhanced,Paris2009QUANTUM,Demkowicz-Dobrzanski2014Quantum,Pezze2014Quantum}. Recent findings show that the quantum Fisher information is the convex roof of the variance, apart from a constant factor \cite{Toth2013Extremal,Yu2013Quantum}, which allowed, for instance, to derive novel uncertainty relations \cite{Toth2022Uncertainty,Chiew2022Improving}, and will also be used in this article.

The paper is organized as follows. In \SEC{sec:QFI}, we summarize basic facts connected to quantum metrology.  In \SEC{sec:Ent}, we summarize entanglement theory. In \SEC{sec:twocopy}, we show how to transform the optimization over decompositions of the density matrix into an optimization of an expectation value over separable states. In \SEC{sec:Wasserstein}, we show some applications of these ideas for the Wasserstein distance. We define a novel type of Wasserstein distance based on an optimization over separable states.    In \SEC{sec:var}, we define variance-like quantities from the Wasserstein distance. In \SEC{sec:wassent}, we discuss how such a Wasserstein distance and the variance-like quantity mentioned above can be used to construct entanglement criteria. In \SEC{sec:variance_instead_of_second_moment}, we introduce further quantities similar to the formulas giving the Wasserstein distance, but they involve the variance of two-body quantities rather than the second moment. In \SEC{sec:other}, we consider an optimization over various subsets of the quantum states. In \SEC{sec:extension}, we extend our ideas to various generalized quantum Fisher information quanitities.

\section{Quantum metrology}
\label{sec:QFI}

Before discussing our main results, we review some of the fundamental relations of quantum metrology. A basic metrological task is estimating the small angle $\theta$ in a unitary dynamics
\begin{equation}
U_{\theta}=\exp(-i{\Hamiltonian}\theta),
\end{equation}
where $\Hamiltonian$ is the Hamiltonian. The precision is limited by the Cram\'er-Rao bound as \cite{Helstrom1976Quantum,Holevo1982Probabilistic,Braunstein1994Statistical,Braunstein1996Generalized,Petz2008Quantum,Giovannetti2004Quantum-Enhanced,Demkowicz-Dobrzanski2014Quantum,Pezze2014Quantum,Toth2014Quantum,Pezze2018Quantum,Paris2009QUANTUM,Barbieri2022Optical}
\begin{equation} \label{eq:cramerrao}
\va{\theta}\ge\frac 1 {\nu {\mathcal F}_Q[\varrho,{\Hamiltonian}]},
\end{equation}
where the factor $1/\nu$ in  \EQ{eq:cramerrao} is the statistical improvement when performing independent measurements on identical copies of the probe state, and the quantum Fisher information is defined by the formula \cite{Helstrom1976Quantum,Holevo1982Probabilistic,Braunstein1994Statistical,Braunstein1996Generalized,Petz2008Quantum}
\begin{equation}
\label{eq:FQ}
{\mathcal F}_Q[\varrho,{\Hamiltonian}]=2\sum_{k,l}\frac{(\lambda_{k}-\lambda_{l})^{2}}{\lambda_{k}+\lambda_{l}}\vert \langle k \vert {\Hamiltonian} \vert l \rangle \vert^{2},
\end{equation}
where the density matrix has the eigendecomposition
\begin{equation}\label{eq:rho_eigdecomp}
\varrho=\sum_{k}\lambda_k \ketbra{k}.
\end{equation}

The quantum Fisher information is bounded from above by the variance and from below by the Wigner-Yanase skew information as
\begin{equation}
4I_{\varrho}(\Hamiltonian)\le {\mathcal F}_Q[\varrho,\Hamiltonian] \leq 4(\Delta \Hamiltonian)^2_{\varrho}\label{eq:FQvar},
\end{equation}
where  if $\varrho$ is pure then the equality holds for both inequalities \cite{Braunstein1994Statistical}. For clarity, we add that the variance is defined as
\begin{equation}
\va{H}_\varrho=\ex{H^2}_\varrho-\ex{H}^2_\varrho,
\end{equation}
where the expectation value is calculated as
\begin{equation}
\ex{H}_\varrho={\rm Tr}(\varrho H).
\end{equation}

The quantum Fisher information is the convex roof of the variance times four  \cite{Toth2013Extremal,Yu2013Quantum,Toth2014Quantum}
\begin{equation}
{\mathcal F}_Q[\varrho,\Hamiltonian]=4\min_{\{p_k,\ket{\Psi_k}\}}\;\sum_k p_k \va{\Hamiltonian}_{\Psi_k},\label{eq:deffqroof}
\end{equation}
where the optimization is carried out over pure state decompositions of the type
\begin{equation}\label{eq:purestatedecomp}
\varrho=\sum_{k}p_k \ketbra{\Psi_k}.
\end{equation}
For the probabilities $p_k\ge0$ and $\sum_k p_k=1$ hold, and the pure states $\ket{\Psi_k}$ are not assumed to be orthogonal to each other.

Apart from the quantum Fisher information, the variance can also be given as a roof \cite{Toth2013Extremal,Yu2013Quantum}
\begin{equation}
\va{\Hamiltonian}_\varrho=\max_{\{p_k,\ket{\Psi_k}\}}\;\sum_k p_k \va{\Hamiltonian}_{\Psi_k}.\label{eq:defvarroof}
\end{equation}

Note that convex and concave roofs of more complicated expressions can also be computed. For instance,
\begin{equation}
\min_{\{p_k,\ket{\Psi_k}\}}  \sum_k p_k \sum_{n=1}^N \va{\Hamiltonian_n}_{\Psi_k}\ge \frac1 4 \sum_{n=1}^N {\mathcal F}_Q[\varrho,\Hamiltonian_n],
\end{equation}
and the expression
\begin{equation}
\max_{\{p_k,\ket{\Psi_k}\}}  \sum_k p_k \sum_{n=1}^N \va{\Hamiltonian_n}_{\Psi_k} \le \sum_{n=1}^N \va{\Hamiltonian_n}_{\varrho},\label{eq:sumvar}
\end{equation}
are the convex and concave roofs, respectively, of the sum of several variances over the decompositions of $\varrho$ given in \EQ{eq:purestatedecomp}. They play a role in the derivation of entanglement conditions \cite{Toth2022Uncertainty}, and will also appear in our results about the quantum Wasserstein distance. We add that there is an equality in \EQ{eq:sumvar} for $N=2$ \cite{Leka2013Some_B,Petz2014}.

Finally, we note that the quantum Fisher information can also be given with a minimization over purifications, which has been used, for instance, to study  quantum metrology in noisy systems \cite{Fujiwara2008A,Escher2011General,Demkowicz-Dobrzanski2012The}. The relation of this finding to the expression in \EQ{eq:deffqroof} is discussed in \REF{Marvian2022Operational}.

\section{Entanglement theory}
\label{sec:Ent}

Next, we review entanglement theory  \cite{Horodecki2009Quantum,Guhne2009Entanglement,Friis2019}. A bipartite quantum state is separable if it can be given as  \cite{Werner1989Quantum}
\begin{equation}\label{eq:sep_states}
\sum_{k}p_{k} \ketbra{\Psi_k}\otimes  \ketbra{\Phi_k},
\end{equation}
where  $p_k$ are for the probabilities, $\ket{\Psi_k}$ and $\ket{\Phi_k}$ are pure quantum states. We will denote the set of separable states by $\mathcal S.$ The mixture of two separable states is also separable, thus the set $\mathcal S$ is convex. If a quantum state cannot be written as \EQ{eq:sep_states} then it is called entangled.

A relevant subset of separable states are the symmetric separable states, which can be given as \cite{Eckert2002Quantum,Ichikawa2008Exchange}
\begin{equation}\label{eq:sym_sep_states}
\sum_{k}p_{k} \ketbra{\Psi_k}\otimes  \ketbra{\Psi_k}.
\end{equation}
We will denote the set of symmetric separable states by $\mathcal S'.$ The mixture of two symmetric separable states is also symmetric and separable, thus the set $S'$ is also convex. Clearly
\begin{equation}
\mathcal S'\subset \mathcal S. \label{eq:SprimeS}
\end{equation}
For such states for the expectation value
\begin{equation}
\ex{P_s}=1\label{eq:Ps1}
\end{equation}
holds, where $P_s$ is the projector to the symmetric subspace defined by the basis vectors $\ket{nn}$ and $(\ket{nm}+\ket{mn})/\sqrt{2}$ for $n\ne m.$
Equivalently, any state in $S'$ fulfills
\begin{equation}
F_{12}\varrho=\varrho F_{12}=\varrho,\label{eq:symprop}
\end{equation}
where $F_{12}$ is the flip operator for which
\begin{equation}
F_{12}\ket{m}\otimes\ket{n}=\ket{n}\otimes\ket{m}\label{eq_flipop}
\end{equation}
for all $m,n.$

The set of quantum states with a positive partial transpose (PPT) $\mathcal P$ consists of the states for which
\begin{equation}
\varrho^{T_k}\ge 0\label{eq:PPTcond}
\end{equation}
holds for $k=1,2,$ where $T_k$ is a partial transposition according to the $k^{th}$ subsystem. $\mathcal P$ is clearly a convex set. Moreover, all separable states given in \EQ{eq:sep_states} fulfill \EQ{eq:PPTcond},
thus
\begin{equation}
\mathcal P\supset \mathcal S
\end{equation}
holds. For two qubits and for a qubit-qutrit system, i.e., for $2\times2$ and $2\times3$ systems the set of PPT states equals the set of separable states \cite{Horodecki1997Separability,Peres1996Separability,Horodecki1999Bound}.

It is generally very difficult to decide whether a state is separable or not, while it is very simple to decide whether the condition given in \EQ{eq:PPTcond} is fulfilled. Such conditions can be even part of semidefinite programs used to solve various optimization problems (e.g., \REF{Toth2018Quantum}). Thus, often the set of PPT quantum states, $\mathcal P$ is used instead of the separable states. Since for small systems $\mathcal P=\mathcal S,$ optimization problems over separable states can be solved exactly for those cases. For larger systems, by optimizing over states in $\mathcal P$ instead of states in $\mathcal S,$ we get lower or upper bounds.

Finally, we will define the set of symmetric PPT states $\mathcal P'.$ States in this set are symmetric, thus \EQS{eq:Ps1} and \eqref{eq:symprop} hold. For the various sets mentioned above, we have the relation
\begin{equation}
\mathcal S' \subset \mathcal P'\subset \mathcal P,
\end{equation}
while for $2\times2$ and $2\times3$ systems, we have $P'=S'.$

Convex roofs and concave roofs play also a central role in entanglement theory. The entanglement of formation is defined as a convex roof over pure components of the von Neumann entropy of the reduced state \cite{Hill1997Entanglement,Wootters1998Entanglement}
\begin{equation}
E_F(\varrho)=\min_{\{p_k,\ket{\Psi_k}\}} \sum_k p_k E_E(\Psi_k),\label{eq:EF}
\end{equation}
where the optimization is over the decompositions given in \EQ{eq:purestatedecomp}, and the entanglement entropy of the pure components is given as
\begin{equation}
E_E(\Psi_k)=S[{\rm Tr}_A(\Psi_k)],
\end{equation}
where $S$ is the von Neumann entropy. $E_F(\varrho)$ is the minimum entanglement needed to create the state. On the other hand, the entanglement of assistance is obtained as a concave roof \cite{DiVincenzo1999Proceedings,Smolin2005Entanglement}
\begin{equation}
E_A(\varrho)=\max_{\{p_k,\ket{\Psi_k}\}}\;\sum_k p_k E_E(\Psi_k).\label{eq:EA}
\end{equation}

The above quantities correspond to the following scenario. Let us assume that the bipartite quantum state $\varrho$ living on parties $A$ and $B$ is realized as the reduced state of a pure state living on parties $A, B$ and $C.$ Let us assume that party $C$ makes a von Neumann measurement resulting in a state $\ket{\Psi_k}$ on $A$ and $B,$ and it sends the measurement result $k$ to $A$ and $B.$ After repeating this on many copies of the three-partite state, the average entanglement of $A$ and $B$ will be
\begin{equation}
\sum_k p_k E_E(\Psi_k),\label{eq:avE}
\end{equation}
where $p_k$ is the probability of the outcome $k$ and  and for the $\ket{\Psi_k}$ states \EQ{eq:purestatedecomp} holds. If party C wants to help the parties $A$ and $B$ to have a large average entanglement, then \EQ{eq:avE} can reach the entanglement of assistance given in \EQ{eq:EA}. On the other hand, the average entanglement is always larger than or equal to the entanglement of formation given in \EQ{eq:EF}.

Next, we will introduce an entanglement condition based on the sum of several variances \cite{Hofmann2003Violation,Guhne2004Characterizing,Guhne2006Entanglement,Vitagliano2011Spin}, which will be used later in the article. Let us consider a full set of traceless observables $\{G_n\}_{n=1}^{d^2-1}$ for $d$-dimensional systems fulfilling
\begin{equation}
{\rm Tr}(G_nG_{n'})=2\delta_{nn'}.
\end{equation}
Any traceless Hermitian observable can be obtained as a linear combination of $G_n.$ In other words, $G_n$ are the SU($d$) generators.
It is known that for pure states  (see e.g., \REF{Vitagliano2011Spin})
\begin{equation}\label{eq:Gkequality}
\sum_{n=1}^{d^2-1} \va{G_n} = 2(d-1)
\end{equation}
holds. Due to the concavity of the variance it follows that for mixed states we have \cite{Vitagliano2011Spin}
\begin{equation}\label{eq:varGk}
\sum_{n=1}^{d^2-1} \va{G_n} \ge 2(d-1).
\end{equation}

Let us now consider a $d\times d$ system. For a product state, $\ket{\Psi}\otimes\ket{\Phi},$ we obtain \cite{Hofmann2003Violation,Vitagliano2011Spin}
\begin{align}
& \sum_{n=1}^{d^2-1} \vasq{(G_n^T \otimes \openone - \openone \otimes G_n)}_{\Psi\otimes\Phi} \nonumber\\
&\quad\quad\quad\quad = \sum_{n=1}^{d^2-1} \va{G_n^T}_{\Psi}+\sum_{n=1}^{d^2-1} \va{G_n}_{\Phi}\nonumber\\
&\quad\quad\quad\quad = 4(d-1),
\label{eq:entond1}
\end{align}
where in the last inequality we used that for pure states \EQ{eq:Gkequality} holds. We also used the fact that if  $\{G_n\}_{n=1}^{d^2-1}$ is a full set of observables with the properties mentioned above, then  $\{G_n^T\}_{n=1}^{d^2-1}$ is also a full set of such observables. Then, for bipartite separable states given in \EQ{eq:sep_states}  \cite{Hofmann2003Violation,Guhne2004Characterizing,Guhne2006Entanglement,Vitagliano2011Spin}
\begin{equation}
\sum_{n=1}^{d^2-1} \vasq{(G_n^T \otimes \openone - \openone \otimes G_n)} \ge 4(d-1)\label{eq:entond}
\end{equation}
holds due to the concavity of the variance. Any state that violates the inequality in \EQ{eq:entond} is entangled.
The left-hand side of \EQ{eq:entond} is zero for the maximally entangled state
\begin{equation}
\ket{\Psi_{\rm me}}=\frac{1}{\sqrt d}\sum_{k=0}^{d-1} \ket{k}\ket{k}.\label{eq:mestate}
\end{equation}
Thus, we say that the criterion given in \EQ{eq:entond} detects entangled states in the vicinity of the state given in \EQ{eq:mestate}.

Let us consider the $d=2$ case concretely. Let us choose $\{G_n\}_{n=1}^3=\{\sigma_x,\sigma_y,\sigma_z\}.$ \EQL{eq:entond} can rewritten as
\begin{align}
&[\Delta (\sigma_x\otimes\openone-\openone\otimes\sigma_x)]^2\nonumber\\
&\quad\quad\quad\quad+
[\Delta (\sigma_y\otimes\openone+\openone\otimes\sigma_y)]^2\nonumber\\
&\quad\quad\quad\quad+
[\Delta (\sigma_z\otimes\openone-\openone\otimes\sigma_z)]^2\ge 4,\label{eq:2qubit}
\end{align}
where $\sigma_l$ are Pauli spin matrices defined as
\begin{align}
\sigma_x&=\left(\begin{array}{cc}0 & 1\\1 & 0\end{array}\right),\quad \sigma_y=\left(\begin{array}{cc}0 & -i\\i & 0\end{array}\right),\nonumber\\
\sigma_z&=\left(\begin{array}{cc}+1 & 0\\0 & -1\end{array}\right).
\end{align}
The left-hand side of the inequality given in \EQ{eq:2qubit} is zero for the state
\begin{equation}
\frac 1 {\sqrt{2}} (\ket{00}+\ket{11}). \label{eq:me_d2}
\end{equation}

We can have a similar construction with only three operators for any system size. Let us consider the usual angular momentum operators $j_x, j_y$ and $j_z,$ living in a $d$-dimensional system, fulfilling
\begin{equation}
j_x^2+j_y^2+j_z^2=j(j+1)\openone,
\end{equation}
where $d=2j+1.$ In particular, let us define the matrix \cite{Edmonds1957Angular}
\begin{equation}
j_z={\rm diag}(-j,-j+1,...,j-1,j).
\end{equation}
Then, we need the ladder operators
\begin{equation}
(j_+)_{m,n}=\delta_{m,n+1} \sqrt{j(j+1)-(j-n)(j+1-n)}\label{eq:jplus}
\end{equation}
and $j_-=j_+^\dagger,$ where $1\le m,n\le d$ and $\delta_{kl}$ is the well-known Kronecker delta. Then, we define the $x$ and $y$ components as
\begin{equation}
j_x=\frac{j_++j_-}{2},\quad\quad j_y=\frac{j_+-j_-}{2i}.
\end{equation}

After clarifying the operators used, we need the uncertainty relation
\begin{equation}
\va{j_x}+\va{j_y}+\va{j_z}\ge j,\label{eq:uncjxyz}
\end{equation}
which is true for all quantum states. Based on \EQ{eq:uncjxyz}, it can be proved that for separable states we have \cite{Toth2004Entanglement,Toth2007Optimal,Toth2010Generation,Vitagliano2011Spin}
\begin{align}
&[\Delta (j_x^T\otimes\openone-\openone\otimes j_x)]^2\nonumber\\
&\quad\quad\quad\quad+
[\Delta (j_y^T\otimes\openone-\openone\otimes j_y)]^2\nonumber\\
&\quad\quad\quad\quad+
[\Delta (j_z^T\otimes\openone-\openone\otimes j_z)]^2\ge 2j.\label{eq:2qubitb}
\end{align}
c.~f.~\EQ{eq:2qubit}. The left-hand side of the inequality given in \EQ{eq:2qubitb} is zero for the maximally entangled state given in \EQ{eq:mestate}. It is easy to see that \EQ{eq:2qubitb} is a tight inequality for separable states, since the \mbox{$\ket{+j,+j}$} state saturates it.

We now present a simple expression for which the maximum  for general states is larger than the maximum for separable quantum states. We know that for separable states \cite{Toth2007Detection,Toth2007Optimal}
\begin{equation}
\ex{(\sigma_x\otimes\openone-\openone\otimes\sigma_x)^2+ (\sigma_y\otimes\openone-\openone\otimes\sigma_y)^2}\le 6,\label{eq:ineqsep}
\end{equation}
while the maximum for quantum states is $8$ and it is taken by the singlet state
\begin{equation}
\frac 1{\sqrt{2}} (\ket{01}-\ket{10}).\label{eq:maxstate}
\end{equation}
We add that the product state $\ket{01}_x\label{eq:01}$ saturates the inequality \EQ{eq:ineqsep}, where $\ket{.}_x$ is a state given in the $x$-basis. That is, for a qubit, the basis states in the $x$-basis are
\begin{align}
\ket{0}_x&=\frac{1}{\sqrt 2}(\ket{0}+\ket{1}),\nonumber\\
\ket{1}_x&=\frac{1}{\sqrt 2}(\ket{0}-\ket{1}).
\end{align}

Let us now determine a complementary relation. We will find the minimum for separable states for the left-hand side of \EQ{eq:ineqsep}. We need to know that for separable states
\begin{equation}
[\Delta (\sigma_x\otimes\openone-\openone\otimes\sigma_x)]^2+[\Delta (\sigma_y\otimes\openone-\openone\otimes\sigma_y)]^2\ge 2\label{eq:2var}
\end{equation}
holds. This can be seen as follows \cite{Hofmann2003Violation,Guhne2004Characterizing}. We need to know that for two-qubit quantum states
\begin{equation}
\va{\sigma_x}+\va{\sigma_y}\ge 2 \label{eq:xy2}
\end{equation}
holds. For product states $\ket{\Psi} \otimes \ket{\Phi},$ the left-hand side of \EQ{eq:2var} equals the sum of single system variances, which can be bounded from below as
\begin{equation}
\va{\sigma_x}_{\Psi}+\va{\sigma_y}_{\Psi}+\va{\sigma_x}_{\Phi}+\va{\sigma_y}_{\Phi} \ge 4. \label{eq:xy4}
\end{equation}
The bound for separable states given in \EQ{eq:sep_states} is the same as the bound for product states, since the variance is a concave function of the state. The left-hand side of \EQ{eq:2var} is zero for the state
\begin{equation}
\frac 1{\sqrt{2}} (\ket{01}+\ket{10}). \label{eq:maxstate2}
\end{equation}

The product state $\ket{11}_x$ saturates the inequality given in \EQ{eq:2var} and for that state
\begin{align}
\ex{\sigma_x\otimes\openone-\openone\otimes\sigma_x}&=0,\nonumber\\
\ex{\sigma_y\otimes\openone-\openone\otimes\sigma_y}&=0
\label{eq:exp0}
\end{align}
hold. Thus, for separable states
\begin{equation}
\ex{(\sigma_x\otimes\openone-\openone\otimes\sigma_x)^2+(\sigma_y\otimes\openone-\openone\otimes\sigma_y)^2}\ge 2\label{eq:2var2}
\end{equation}
holds and it is even a tight inequality for separable states, c. f. \EQ{eq:ineqsep}.

\section{Optimization over the two-copy space}
\label{sec:twocopy}

In this section, we first review the formalism that maps the optimization over the decompositions of the density matrix to an optimization of an operator expectation value over bipartite symmetric separable quantum states with given marginals \cite{Toth2015Evaluating} \footnote{For a reference in the literature, see Eq.~(67) in \REF{Toth2013Extremal} and Eqs.~(S23)-(S25) in \REF{Toth2015Evaluating}. In Eqs.~(S23)-(S25) in \REF{Toth2015Evaluating}, there are some additional constraints for the expectation values $\exs{O_i}_\varrho,$ where $O_i$ are some operators. For our discussion, they are not needed.}. Then, we show that the same result is obtained if the optimization is carried out over general separable quantum states rather than symmetric separable states.

We start from writing the variance of a pure  state $\ket{\Psi}$ as an operator expectation value acting on two copies as
\begin{equation}
\va{\Hamiltonian}_\Psi=\trace(\Omega\ketbra{\Psi}\otimes \ketbra{\Psi}),\label{eq:twocopy}
\end{equation}
where we define the operator
\begin{equation}
\Omega=\Hamiltonian^2\otimes \openone-\Hamiltonian\otimes \Hamiltonian.
\end{equation}
Based on \EQ{eq:twocopy}, the expression of the quantum Fisher information given in \EQ{eq:deffqroof} can be rewritten as \cite{Toth2013Extremal,Toth2015Evaluating}
\begin{align}
{\mathcal F}_Q[\varrho,\Hamiltonian]=\nonumber\\\nonumber
4\min_{\{p_k,\ket{\Psi_k}\}} \;&
\sum_k p_k \trace[\Omega(\ketbra{\Psi_k})^{\otimes 2}],\\
\textrm{s.~t. }&
\varrho=\sum_{k}p_k \ketbra{\Psi_k}.
\end{align}
The sum can be moved into the trace and we obtain
\begin{align}
{\mathcal F}_Q[\varrho,\Hamiltonian]=\nonumber\\\nonumber
4\min_{\{p_k,\ket{\Psi_k}\}} \;&
 \trace\left\{\Omega \left[\sum_k p_k(\ketbra{\Psi_k})^{\otimes 2}\right]\right\},\\
\textrm{s.~t. }&
\varrho=\sum_{k}p_k \ketbra{\Psi_k}. \label{eq:squarebracket}
\end{align}
On the right-hand side in \EQ{eq:squarebracket} in the square bracket, we can recognize a mixed state living in the two-copy space
\begin{equation}\label{omega12}
\varrho_{12}=\sum_{k}p_{k} \ketbra{\Psi_k}\otimes  \ketbra{\Psi_k},
\end{equation}
States given in \EQ{omega12} are symmetric separable states, thus $\varrho_{12}\in\mathcal S'.$ We can rewrite the optimization in \EQ{eq:squarebracket} as an optmization over symmetric separable states given in \EQ{omega12} as
\begin{align}
{\mathcal F}_Q[\varrho,\Hamiltonian]=
\min_{ \varrho_{12} }\;&
4\trace(\Omega\varrho_{12} ),\nonumber\\
\textrm{s.~t. }&
\varrho_{12} \in \mathcal S',\nonumber\\
& {\rm Tr}_2(\varrho_{12})=\varrho.\label{eq:deffqroof2}
\end{align}
Due to the optimization over symmetric separable states, ${\rm Tr}_1(\varrho_{12})=\varrho$ is fulfilled without adding it as an explicit constraint.

\EQL{eq:deffqroof2}  contains an optimization over symmetric separable states, which cannot be computed directly. However, we can consider an optimization over a larger set, the set of symmetric PPT states, which leads to an expression that can be obtained numerically using semidefinite programming  \cite{Vandenberghe1996Semidefinite}
\begin{align}
{\mathcal F}_Q^{({\rm PPT})}[\varrho,\Hamiltonian]=
\min_{ \varrho_{12} }\;&
4\trace(\Omega\varrho_{12} ),\nonumber\\
\textrm{s.~t. }&
\varrho_{12}\in\mathcal P',\nonumber\\
& {\rm Tr}_2(\varrho_{12})=\varrho.
\label{Fqopt}
\end{align}
In general, the relation
\begin{equation}
{\mathcal F}_Q^{({\rm PPT})}[\varrho,\Hamiltonian]\le {\mathcal F}_Q[\varrho,\Hamiltonian]
\end{equation}
holds, while for two qubits we have an equality, since for that system size $\mathcal S'=\mathcal P',$ as we discussed in \SEC{sec:Ent}.

Let us now change the operator to be optimized, making it permutationally symmetric. \EQL{eq:deffqroof2} can be rewritten as
\begin{align}
{\mathcal F}_Q[\varrho,\Hamiltonian]=\nonumber
\min_{ \varrho_{12} }\;&
2\trace[(\Hamiltonian\otimes \openone-\openone\otimes \Hamiltonian)^2  \varrho_{12} ],\\
\textrm{s.~t. }&
\varrho_{12}\in\mathcal S',\label{eq:deffqroof3}  \nonumber\\
& {\rm Tr}_2(\varrho_{12})=\varrho .
\end{align}

We will now show that the expression in \EQ{eq:deffqroof3}  remains true if we change the set over which we have to optimize to the set of separable states.

\DEFOBS{obs:QFI_sep}The quantum Fisher information can be obtained as an optimization over separable states as
\begin{align}
{\mathcal F}_Q[\varrho,\Hamiltonian]=
\min_{ \varrho_{12} }\;&
2\trace[(\Hamiltonian\otimes \openone-\openone\otimes \Hamiltonian)^2  \varrho_{12} ],\nonumber\\
\textrm{s.~t. }&
\varrho_{12}\in \mathcal S,\nonumber\\
& {\rm Tr}_2(\varrho_{12})=\varrho,\nonumber\\
& {\rm Tr}_1(\varrho_{12})=\varrho.\label{eq:FQ_min_sep_states}
\end{align}

\PROOF Using
\begin{align}
&\trace[(\Hamiltonian\otimes \openone-\openone\otimes \Hamiltonian)^2  \varrho_{12} ]\nonumber\\
&\quad\quad=2\trace(\Hamiltonian^2 \varrho)-2\trace[(\Hamiltonian\otimes \Hamiltonian)\varrho_{12}],
\end{align}
the right-hand side of \EQ{eq:FQ_min_sep_states} can be reformulated  as
\begin{align}
4\trace(\Hamiltonian^2 \varrho)-4\max_{\varrho_{12}}\;&
\trace[(\Hamiltonian\otimes \Hamiltonian)  \varrho_{12} ],\nonumber\\
\textrm{s.~t. }&
\varrho_{12}\in\mathcal S,\label{eq:deffqroof4}  \nonumber\\
& {\rm Tr}_2(\varrho_{12})=\varrho,\nonumber\\
& {\rm Tr}_1(\varrho_{12})=\varrho.
\end{align}
Due to the permutational invariance of $\Hamiltonian\otimes \Hamiltonian$ and that both marginals must be equal to $\varrho,$ if a separable state given in \EQ{eq:sep_states} maximizes the correlation $\ex{\Hamiltonian\otimes \Hamiltonian}$, then the separable state
\begin{equation}\label{sep2}
\sum_{k}p_{k} \ketbra{\Phi_k}\otimes  \ketbra{\Psi_k}.
\end{equation}
also maximizes the correlation. Then, the mixture of the above two separable states also maximize the correlation with the given marginals, which can always be written as
\begin{equation}
\varrho_{12}= \sum_{k=1}^M \tilde p_k \ketbra{\tilde \Psi_k} \otimes  \ketbra{\tilde \Psi_{\pi_k}},\label{sep3}
\end{equation}
where $\pi_k$ for $k=1,2,..,M$ is a permutation of $1,2,...,M$ for some $M.$  Based on our arguments, it is sufficient to look for the separable state that maximizes $\ex{\Hamiltonian\otimes \Hamiltonian}$ in the form \EQ{sep3}, and $\tilde p_k\ge0,$ $\sum_k \tilde p_k=1.$ Then, the correlation can be given as
\begin{equation}
\trace[(\Hamiltonian\otimes \Hamiltonian)  \varrho_{12} ]= \sum_{k=1}^M \tilde p_k h_k h_{\pi_k},
\end{equation}
where we define the expectation values of $\Hamiltonian$ on $\ket{\tilde\Psi_k}$ as
\begin{equation}
h_k=\trace(\Hamiltonian \ketbra{\tilde\Psi_k}).
\end{equation}
Based on the  Cauchy-Schwarz  inequality we have
\begin{align}
\sum_k \tilde p_k h_k h_{\pi_k} &\le \sqrt{\sum_k \tilde p_k h_k^2\sum_k \tilde p_k h_{\pi_k}^2}\nonumber\\
&\le \max\left(\sum_k \tilde p_k h_k^2,\sum_k \tilde p_k h_{\pi_k}^2\right).\label{eq:ineqCS}
\end{align}
Thus, when we maximize $\trace[(\Hamiltonian\otimes \Hamiltonian)  \varrho_{12} ]$ over separable states with the constraints for the marginals, the maximum is taken by a symmetric separable state given in \EQ{omega12}. In this case both inequalities are saturated in \EQ{eq:ineqCS}. $\qed$

Based on \EQ{eq:defvarroof}, we can obtain the variance also as a result of an optimization over a two-copy space.

\DEFOBS{obs:VAR_sep}The variance can also be obtained as an optimization over symmetric separable states as
\begin{align}
\va{\Hamiltonian}=
\max_{ \varrho_{12} }\;&
\frac1 2 \trace[(\Hamiltonian\otimes \openone-\openone\otimes \Hamiltonian)^2  \varrho_{12} ],\nonumber\\
\textrm{s.~t. }&
\varrho_{12}\in \mathcal S',\nonumber\\
& {\rm Tr}_2(\varrho_{12})=\varrho,\nonumber\\
& {\rm Tr}_1(\varrho_{12})=\varrho.\label{eq:VAR_max_sep_states}
\end{align}

\DEFEXAMPLE{ex:ex0}Note that if we replace the set of symmetric separable states by the set of separable states in \EQ{eq:VAR_max_sep_states}, then we get a different quantity, which can be larger in some cases than $\va{\Hamiltonian}.$ Let us see a concrete example. For instance, if $\varrho=\openone/2$ then among symmetric separable states, the maximum is $1$ and it is attained by the state
\begin{equation}
\varrho_{12}=\frac1 4\left(\ketbra{00}+\ketbra{11}+2\ketbra{\Psi^+}\right),\label{eq:symsep}
\end{equation}
where the Bell state $\ket{\Psi^+}$ is given as
\begin{equation}
\ket{\Psi^+}=\frac1{\sqrt 2}(\ket{01}+\ket{10}).
\end{equation}
The state given in \EQ{eq:symsep} can be decomposed into the mixture of symmetric product states as
\begin{equation}
\frac1 4(\ketbra{\alpha_{+1}}+\ketbra{\alpha_{-1}}+\ketbra{\alpha_{+i}}+\ketbra{\alpha_{-i}}),
\end{equation}
where the symmetric product state is defined as
\begin{equation}
\ket{\alpha_{q}}=(\ket{0}+q\ket{1})\otimes(\ket{0}+q\ket{1}).
\end{equation}
Among separable states, the maximum is $2$ and it is attained by
the state
\begin{equation}
\varrho_{12}=\frac1 2\left(\ketbra{01}+\ketbra{10}\right).\label{eq:rho12sep}
\end{equation}
Note that $\varrho_{12}$ given in \EQ{eq:rho12sep} is not symmetric.

\section{Quantum Wasserstein distance based on a separable coupling}
\label{sec:Wasserstein}

Next, we will connect our result in \OBS{obs:QFI_sep} to results available in the literature mentioned in the introduction. We will consider the quantum Wasserstein distance such that the optimization takes place over separable states rather than over general bipartite quantum states. We will consider such modifications of $D_{\rm GMPC}(\varrho,\sigma)^2$ and $D_{\rm DPT}(\varrho,\sigma)^2,$ and examine their properties.

Our first finding concerning the GMPC distance is the following.

\DEFDEFINITION{def:def1}Modifying the definition in \EQ{eq:GMPC_distance} we can define a new type of GMPC distance such that we restrict the optimization over separable states as
\begin{align}
D_{\rm GMPC,sep}(\varrho,\sigma)^2\quad\nonumber\\=\frac 1 2
\min_{ \varrho_{12}}\sum_{n=1}^N\;&
\trace[(\Hamiltonian_n\otimes \openone-\openone\otimes \Hamiltonian_n)^2  \varrho_{12} ],\nonumber\\
\textrm{s.~t. }&
\varrho_{12}\in\mathcal S,\label{eq:GMPC_distance_sep}  \nonumber\\
& {\rm Tr}_1(\varrho_{12})=\varrho, \nonumber\\
& {\rm Tr}_2(\varrho_{12})=\sigma.
\end{align}

We can see immediately two relevant properties of the newly defined distance. \OBS{obs:QFI_sep} showed that for $N=1$ for a given $H_1$
\begin{equation}
D_{\rm GMPC,sep}(\varrho,\varrho)^2=\frac1 4 {\mathcal F}_Q[\varrho,\Hamiltonian_1]\label{eq:self_distance_GMPC}
\end{equation}
holds. Moreover, based on \EQS{eq:GMPC_distance} and \eqref{eq:GMPC_distance_sep}, we can immediately see that
\begin{equation}
D_{\rm GMPC}(\varrho,\sigma)^2\le D_{\rm GMPC,sep}(\varrho,\sigma)^2, \label{eq:DGMPC2DGMPCsep2}
\end{equation}
since on the left-hand side of \EQ{eq:DGMPC2DGMPCsep2} there is a minimization over a larger set of quantum states than on the right-hand side. Based on \EQS{eq:self_distance_GMPC} and \eqref{eq:DGMPC2DGMPCsep2}, it follows that
for the self-distance for $N=1$ for a given $H_1$ we obtain
\begin{equation}
D_{\rm GMPC}(\varrho,\varrho)^2\le\frac1 4 {\mathcal F}_Q[\varrho,\Hamiltonian_1].
\end{equation}

Now, our goal is to give an optimal transport map (plan) corresponding to an optimal coupling of $D_{GMPC, \rm sep}(\varrho,\sigma)^2.$ Note that there can be several optimal couplings. We look for  a CPTP map $\Phi$ corresponding to an optimal separable coupling $\varrho_{12}.$ Let us assume that an optimal separable coupling equals the separable state given in \EQ{eq:sep_states}. Based on \DEFINITION{def:def1}, it has the following marginals. We obtain $\varrho$  as in \EQ{eq:purestatedecomp}, and for the state $\sigma$
\begin{equation}
\sigma=\sum_k p_k \ketbra{\Phi_k},\label{eq:cond_sigma222}
\end{equation}
holds. Let us consider the map \cite{DePalma2021Quantum}
\begin{equation}\label{eq:map}
\Phi(X)=\sum_kB_kXB_k^{\dagger}.
\end{equation}
where the Kraus operators are given as
\begin{equation}\label{eq:Kraus-op}
B_k=\sqrt{p_k}A_k\varrho^{-1/2},
\end{equation}
and $\varrho^{-1}$ is the inverse of $\varrho$ on its support.

For our particular transport problem, let us choose
\begin{equation}
A_k=\ket{\Phi_k}\bra{\Psi_k}.\label{eq:Ak}
\end{equation}
It is clear that $\Phi$ is completely positive and it is trace preserving, since
\begin{align}
    \sum_kB_k^{\dagger}B_k
    &=\sum_kp_k\varrho^{-1/2}\ket{\Psi_k}\bra{\Phi_k}\Phi_k\rangle\bra{\Psi_k}\varrho^{-1/2}\nonumber\\
    &=\varrho^{-1/2}\sum_kp_k\ketbra{\Psi_k}\varrho^{-1/2}\nonumber\\
    &=\varrho^{-1/2}\varrho\varrho^{-1/2}=\openone.
\end{align}
Since the map transforms $\varrho$ to $\sigma$
\begin{align}
    \Phi(\varrho)&=\sum_k\sqrt{p_k}\ket{\Phi_k}\bra{\Psi_k}\varrho^{-1/2}\varrho\varrho^{-1/2}\ket{\Psi_k}\bra{\Phi_k}\sqrt{p_k}\nonumber\\
    &=\sum_kp_k\ketbra{\Phi_k}=\sigma,\label{eq:transportmap}
\end{align}
the CPTP map $\Phi$ given in Eq.~(\ref{eq:map}) gives the optimal transport map we were looking for.

Let us see some properties of the map we have just found. If $\ket{\Psi_k}$ are pairwise orthogonal to each other then
\begin{equation}
B_k=A_k=\ket{\Phi_k}\bra{\Psi_k}
\end{equation}
holds. In this case, the map can be realized by a von Neumann measurement in the basis given by $\{\ket{\Psi_k}\},$ with a subsequent unitary that transforms $\ket{\Psi_k}$ to $\ket{\Phi_k}.$ It is instructive to look at the action of the map on the state
\begin{equation}
\varrho_0=\sum_k p_k \ketbra{\Psi_k}_1 \otimes \ketbra{\Psi_k}_2.\label{eq:varrho0}
\end{equation}
Then, we obtain the optimal coupling $\varrho_{12}$ as
\begin{equation}
(\openone \otimes \Phi)(\varrho_0)=\sum_k p_k \ketbra{\Psi_k}_1 \otimes \ketbra{\Phi_k}_2.
\end{equation}
Hence, for every map of given by \EQS{eq:map}, \eqref{eq:Kraus-op} and \eqref{eq:Ak} there is a corresponding coupling.

In summary, if we restrict the optimization for separable states, then when computing $D_{\rm GMPC, sep}(\varrho,\sigma)^2,$ for all $\varrho$ and $\sigma,$ there is a transport map corresponding to all optimal couplings. Note that this was not the case for $D_{\rm GMPC}(\varrho,\sigma)^2.$

Let us now define another distance based on an optimization over separable states.

\DEFDEFINITION{def:def2}Based on the definition of $D_{\rm DPT}(\varrho,\sigma)^2$ given in \EQ{eq:distance}, we can also define
\begin{align}
D_{\rm DPT, sep}(\varrho,\sigma)^2\quad\quad\quad\nonumber\\
=\frac 1 2
\min_{ \varrho_{12}}\sum_{n=1}^N\;&
\trace[(\Hamiltonian_n^T\otimes \openone-\openone\otimes \Hamiltonian_n)^2  \varrho_{12} ],\nonumber\\
\textrm{s.~t. }&
\varrho_{12}\in\mathcal S,\label{eq:distance_sep}  \nonumber\\
& {\rm Tr}_2(\varrho_{12})=\varrho^T, \nonumber\\
& {\rm Tr}_1(\varrho_{12})=\sigma.
\end{align}

As an important property of $D_{\rm DPT, sep}(\varrho,\sigma)^2,$ we can see that
\begin{equation}
D_{\rm DPT}(\varrho,\sigma)^2\le D_{\rm DPT, sep}(\varrho,\sigma)^2 \label{eq:D2Dsep2}
\end{equation}
holds, since on the left-hand side of \EQ{eq:D2Dsep2} there is a minimization over a larger set of quantum states than on the right-hand side.

Let us see now, what kind of map corresponds to an optimal separable coupling $\varrho_{12},$ when we calculate $D_{\rm DPT, sep}(\varrho,\sigma)^2.$ Let us assume that an optimal separable coupling is of the form
\begin{equation}
\varrho_{12}=\sum_k p_k \ketbra{\Psi_k^*}_1 \otimes \ketbra{\Phi_k}_2.
\end{equation}
Based on \DEFINITION{def:def2}, we obtain $\varrho$  as in \EQ{eq:purestatedecomp}, and $\sigma$ as in \EQ{eq:cond_sigma222}. It turns out that the map we need is just $\Phi(\varrho)$ defined in \EQ{eq:map}. It is instructive to look at the action of the map on the state
\begin{equation}
\varrho_0^{T_1}=\sum_k p_k \ketbra{\Psi_k^*}_1 \otimes \ketbra{\Psi_k}_2,
\end{equation}
where $\varrho_0$ is defined in \EQ{eq:varrho0}. Then, we obtain the optimal coupling $\varrho_{12}$ as
\begin{equation}
(\openone \otimes \Phi)(\varrho_0^{T_1})=\sum_k p_k \ketbra{\Psi_k^*}_1 \otimes \ketbra{\Phi_k}_2.
\end{equation}
Hence, for every map of given by \EQS{eq:map}, \eqref{eq:Kraus-op} and \eqref{eq:Ak} there is a corresponding coupling.

It is instructive to relate our results to those of \REF{DePalma2021Quantum}. In \REF{DePalma2021Quantum}, the role of $\varrho_0^{T_1}$ is played by the purification.
Moreover, if we compute the self-distance and thus $\varrho=\sigma,$ the map given in \EQ{eq:map} is not the identity map. On the other hand, in \REF{DePalma2021Quantum} the map was the identity map in that case.

Interestingly, the two different distance measures, defined with the transpose and without it, respectively, are equal to each other.

\DEFOBS{obs:distance_sep_equal}The two quantum Wasserstein distance measures are equal to each other
\begin{equation}\label{eq:distance_sep_ineq}
D_{\rm DPT, sep}(\varrho,\sigma)^2=D_{\rm GMPC, sep}(\varrho,\sigma)^2.
\end{equation}

 \PROOF We need to know that for two matrices $X$ and $Y$ acting on a bipartite system
\begin{equation}
{\rm Tr}(XY)={\rm Tr}(X^{T_k}Y^{T_k})
\end{equation}
holds for $k=1,2.$ Then, from \EQ{eq:GMPC_distance_sep} it follows that
\begin{align}
D_{\rm GMPC,sep}(\varrho,\sigma)^2\quad\quad\quad\nonumber\\=\frac 1 2
\min_{ \varrho_{12}}\sum_{n=1}^N\;&
\trace[(\Hamiltonian_n^T\otimes \openone-\openone\otimes \Hamiltonian_n)^2  \varrho_{12}^{T_1} ],\nonumber\\
\textrm{s.~t. }&
\varrho_{12}^{T1}\in\mathcal S, \nonumber\\
& {\rm Tr}_2(\varrho_{12})=\varrho^T, \nonumber\\
& {\rm Tr}_1(\varrho_{12})=\sigma.
\end{align}
Then, we arrive at \EQ{eq:distance_sep} by noticing that
\begin{equation}
\varrho_{12}^{T_1} \in \mathcal S
\end{equation}
holds if and only if
\begin{equation}
\varrho_{12}\in \mathcal S
\end{equation}
holds. $\qed$

It is instructive to obtain a quantum state that maximizes $D_{\rm DPT, sep}(\varrho,\varrho)^2$ and $D_{\rm DPT}(\varrho,\varrho)^2$ for $N=1$ for a given $\Hamiltonian_1$ as follows. We know that
\begin{equation}
I_{\varrho}(\Hamiltonian_1) \leq {\mathcal F}_Q[\varrho,\Hamiltonian_1]/4 \leq (\Delta \Hamiltonian_1)^2_{\varrho} \le \frac1 4(h_{\max}-h_{\min})^2\label{eq:ineqIvar2}
\end{equation}
holds. The first two inequalities are based on \EQ{eq:FQvar}.  The third one can be obtained as follows. Simple algebra shows that for the state
\begin{equation}
\ket{\Psi_{\rm opt}}=\frac1{\sqrt{2}}(\ket{h_{\min}}+\ket{h_{\max}}),\label{eq:Psi_optim}
\end{equation}
where $\ket{h_{\min}}$ ($\ket{h_{\max}}$) is the eigenstate of $\Hamiltonian_1$ with the minimal eigenvalue $h_{\min}$ (maximal eigenvalue $h_{\max}$)  of $\Hamiltonian_1,$ the variance $(\Delta \Hamiltonian_1)^2$ is maximal, and all inequalities of \EQ{eq:ineqIvar2} are saturated. Hence, the maximal self-distance is achieved by $\ket{\Psi_{\rm opt}}$
\begin{align}
&D_{\rm DPT, sep}(\ketbra{\Psi_{\rm opt}},\ketbra{\Psi_{\rm opt}})^2\nonumber\\
&\quad\quad=D_{\rm DPT}(\ketbra{\Psi_{\rm opt}},\ketbra{\Psi_{\rm opt}})^2\nonumber\\
&\quad\quad=\frac1 4(h_{\max}-h_{\min})^2.\label{eq:Doptim2_rho_rho}
\end{align}

Let us calculate $D_{\rm GMPC,sep}(\varrho,\sigma)^2$ and the other quantum Wasserstein distance measures for some concrete examples.

\begin{figure}[t!]
\includegraphics[width=\columnwidth]
{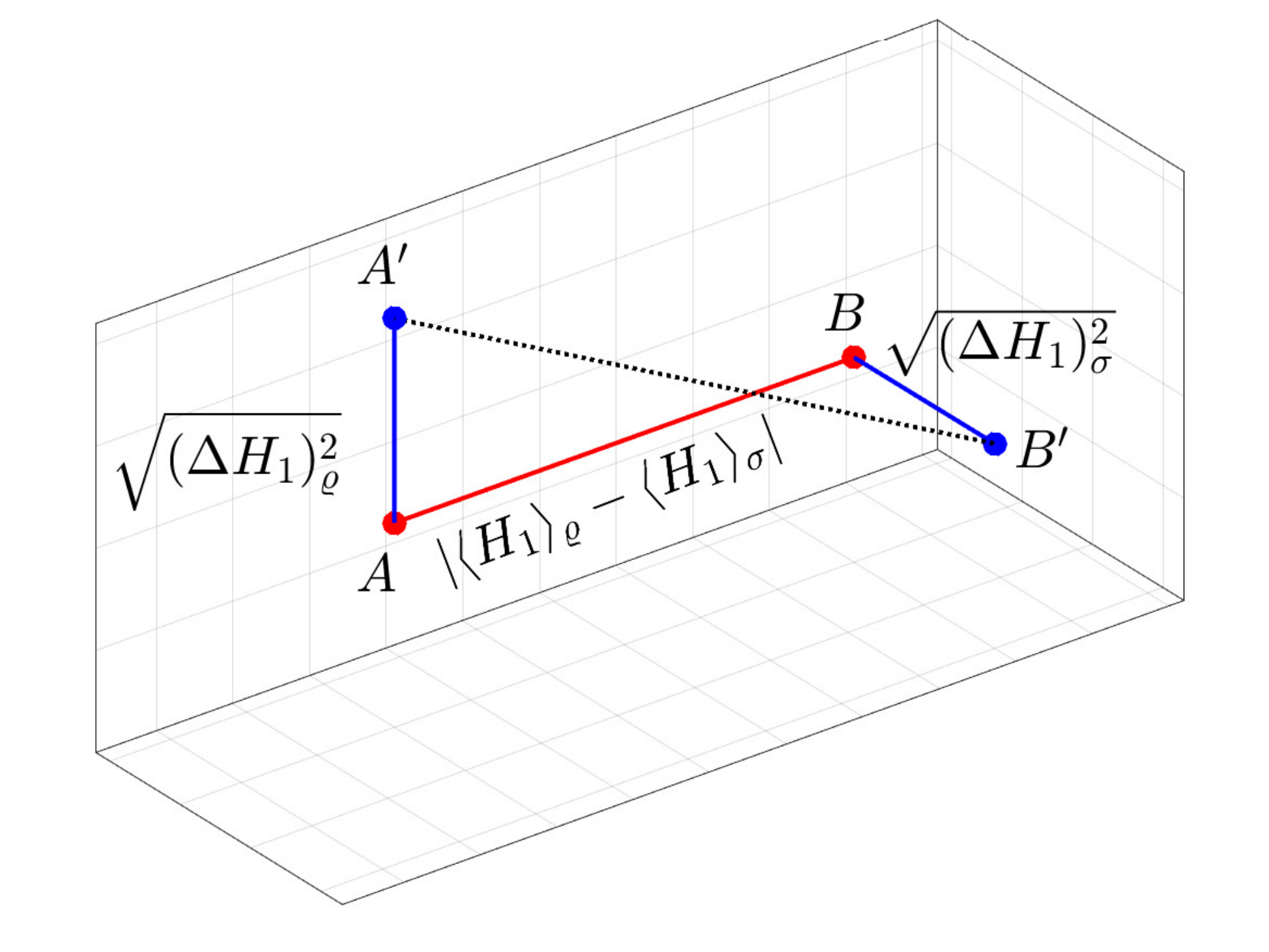}
\caption{Geometric representation of the quantum Wasserstein distance between a pure state $\varrho$ and a mixed state $\sigma$ given in  \EXAMPLE{ex:ex1} for $N=1$ with operator $H_1$. In \EQ{eq:eq:D2_pure_mixedB}, the Wasserstein distance square is $1/2$ times the sum of three terms, corresponding to the two uncertainties and $(\ex{H_1}_{\varrho}-\ex{H_1}_{\sigma})^2.$ Thus, not only the $AB$ distance matters, but the variances computed for the two states. The relevant points are $A(\ex{H_1}_{\varrho},0,0)$, $B(\ex{H_1}_{\sigma},0,0)$, $A'(\ex{H_1}_{\varrho},\sqrt{\va{H_1}_{\varrho}},0)$ and $B'(\ex{H_1}_{\sigma},0,\sqrt{\va{H_1}_{\sigma}}).$ The quantum Wasserstein distance equals $1/\sqrt2$ times the usual Euclidean distance between $A'$ and $B'.$} \label{fig:1D_example}
\end{figure}

\DEFEXAMPLE{ex:ex1}Let us consider the case when $\varrho=\ketbra{\Psi}$ is a pure state of any dimension and $\sigma$ is an arbitrary density matrix of the same dimension. Then, when computing the various Wasserstein distance measures between $\varrho$ and $\sigma,$ the state $\varrho_{12}$ in the optimization is constrained to be the tensor product of the two density matrices. Hence, for the distance from a pure state,
\begin{align}
&D_{\rm DPT}(\ketbra{\Psi},\sigma)^2\nonumber\\
&\quad\quad\quad=D_{\rm DPT, sep}(\ketbra{\Psi},\sigma)^2\nonumber\\
&\quad\quad\quad=D_{\rm GMPC, sep}(\ketbra{\Psi},\sigma)^2\nonumber\\
&\quad\quad\quad=D_{\rm GMPC}(\ketbra{\Psi},\sigma)^2\nonumber\\
&\quad\quad\quad=\sum_{n=1}^N \frac{\ex{\Hamiltonian_n^2}_{\varrho}+\ex{\Hamiltonian_n^2}_{\sigma}}2-\ex{\Hamiltonian_n}_{\varrho}\ex{\Hamiltonian_n}_{\sigma}\nonumber\\
\label{eq:D2_pure_mixed}
\end{align}
holds. The last expression in \EQ{eq:D2_pure_mixed} can be written also as
\begin{equation}
\frac1 2 \sum_{n=1}^N\left[\va{\Hamiltonian_n}_{\varrho}+\va{\Hamiltonian_n}_{\sigma}+(\ex{\Hamiltonian_n}_{\varrho}-\ex{\Hamiltonian_n}_{\sigma})^2\right],
\label{eq:eq:D2_pure_mixedB}
\end{equation}
see \FIG{fig:1D_example}.

\DEFEXAMPLE{ex:ex2}Let us consider the single-qubit states
\begin{equation}
\varrho=\sigma=\frac1 2 \openone.\label{eq:totallymixedrhosigma}
\end{equation}
Let us take $N=1$ and $\Hamiltonian_1=\sigma_z.$ Then, when computing $D_{\rm GMPC,sep}(\varrho,\sigma)^2,$ the optimum is attained by the separable state
\begin{equation}
\varrho_{12}=\frac1 2\ketbra{00} + \frac1 2 \ketbra{11}, \label{eq:rho12_optim}
\end{equation}
and for the self-distance we have
\begin{equation}
D_{\rm GMPC,sep}\left(\frac1 2 \openone,\frac1 2 \openone\right)^2=0
\end{equation}
holds. If $\Hamiltonian_1$ is a different operator then the state $\varrho_{12}$ corresponding to the minimum will be different. It can be obtained from the state given in \EQ{eq:rho12_optim} with local unitaries. For instance, for $\Hamiltonian_1=\sigma_x,$ the optimum is attained by the separable state
\begin{equation}
\varrho_{12}=\frac1 2 \ketbra{00}_x + \frac1 2 \ketbra{11}_x. \label{eq:rho12_optim_x}
\end{equation}

\DEFEXAMPLE{ex:ex3}Let us consider the two-qubit states $\varrho=\sigma=D_p,$ where the diagonal state is defined as
\begin{equation}
D_p=\left(\begin{array}{cc}\phantom{1}p\phantom{-}&0\\\phantom{1}0\phantom{-}&1-p\end{array}\right).
\end{equation}
Let us take $N=1$ and $\Hamiltonian_1=\sigma_z.$ When computing $D_{\rm GMPC,sep}(\varrho,\sigma)^2,$ he optimum is reached by the bipartite separable state
\begin{equation}
\varrho_{12}=p \ketbra{00} + (1-p) \ketbra{11}. \label{eq:rho12_optim_B}
\end{equation}
The self-distance is zero for all $p$
\begin{equation}
D_{\rm GMPC,sep}(D_p,D_p)^2=0.
\end{equation}

\begin{figure}[t!]
\includegraphics[width=0.9\columnwidth]{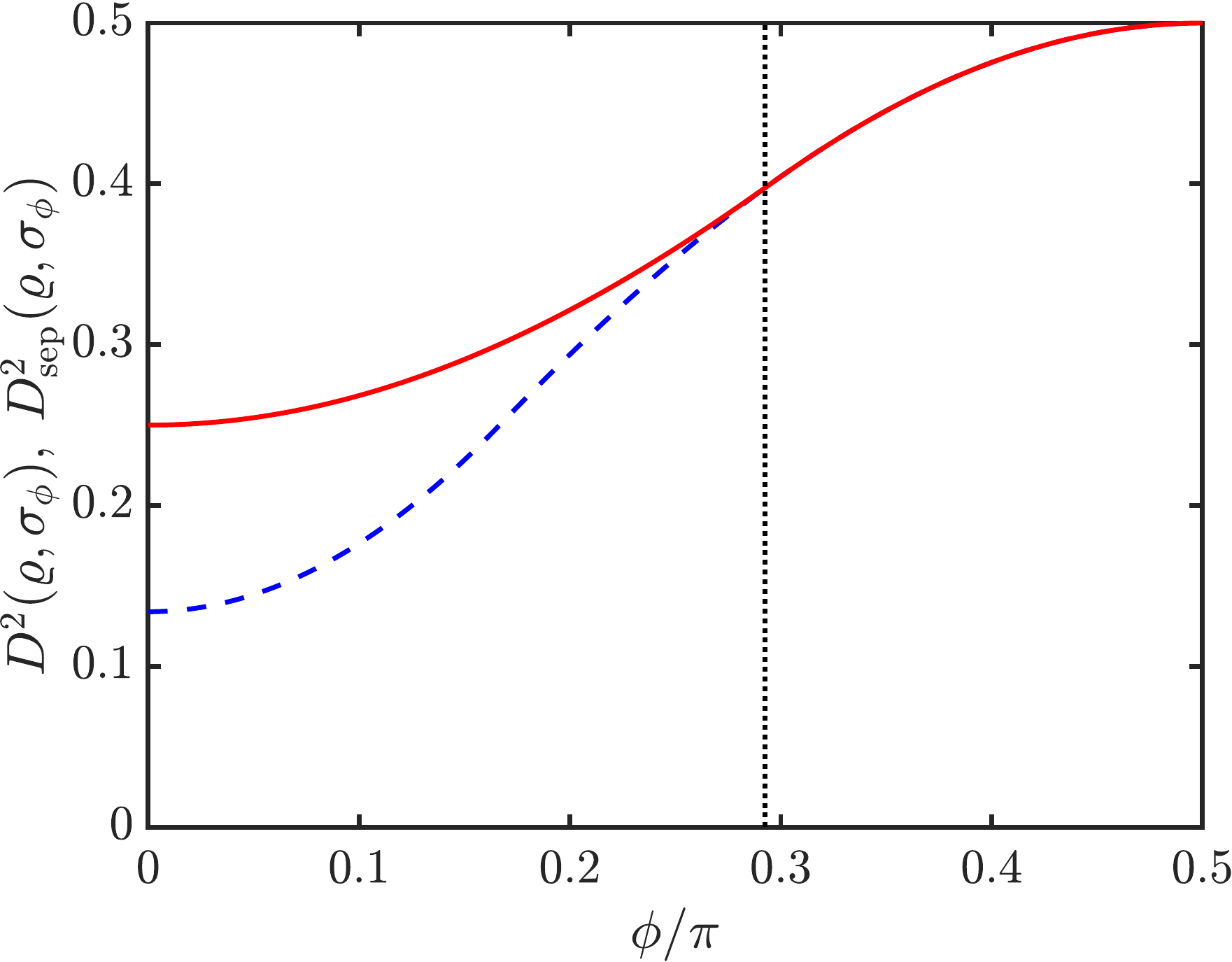}
\caption{(solid) $D_{\rm DPT, sep}(\varrho,\sigma_{\phi})^2$ and (dashed) $D_{\rm DPT}(\varrho,\sigma_{\phi})^2$ for the states given in \EQS{eq:examplestate1} and \eqref{eq:examplestate2}. The two curves coincide with each other on the right-hand side of the vertical dotted line, where $\phi\ge \phi_0$ and $\phi_0$ is given in \EQ{eq:phi0}.} \label{fig:D2sep}
\end{figure}

\DEFEXAMPLE{ex:ex4}Let us consider the two single-qubit mixed states
\begin{equation}
\varrho=\frac1 2\ketbra{1}_x+\frac 1 2 \cdot \frac{\openone}{2},\label{eq:examplestate1}
\end{equation}
and
\begin{equation}
\sigma_{\phi}=e^{-i\frac{\sigma_y}{2}\phi} \varrho e^{+i\frac{\sigma_y}{2}\phi},\label{eq:examplestate2}
\end{equation}
$\phi$ is a real parameter. For $N=1, \Hamiltonian_1=\sigma_z$ we plotted $D_{\rm DPT, sep}(\varrho,\sigma_{\phi})^2$ and $D_{\rm DPT}(\varrho,\sigma_{\phi})^2$ in \FIG{fig:D2sep}. The details of the numerical calculations are in \APP{app:Numerics}. For $\phi=0,$  $\sigma_{\phi}=\varrho,$ hence it this case the two types of distance equal the corresponding types of self-distance of $\varrho.$ That is, based on \EQ{eq:DrhoI}, we have
\begin{equation}
D_{\rm DPT}(\varrho,\varrho)^2=I_\varrho(H_1) = 1-\frac{\sqrt{3}}{2}\approx 0.1340,
\end{equation}
and based on \EQ{eq:self_distance_GMPC} and taking into account \OBS{obs:distance_sep_equal}, we have
\begin{equation}
D_{\rm DPT, sep}(\varrho,\varrho)^2=\frac{{\mathcal F}_Q[\varrho,H_1]}{4}=\frac1 4.
\end{equation}
Here, we used the formula giving the quantum Fisher information with the variance for pure states mixed with white noise as \cite{Toth2012Multipartite,Hyllus2012Fisher,Toth2014Quantum,Toth2020Activating}
\begin{align}
&{\mathcal F}_Q\left[p\ketbra{\Psi}+(1-p)\frac{\openone}d,H\right]\nonumber\\
&\quad\quad\quad\quad\quad\quad=\frac{p^2}{p+2(1-p)d^{-1}} 4 \va{H}_{\Psi},
\end{align}
where $d$ is the dimension of the system. For $\phi=\pi/2,$
\begin{equation}
\sigma_{\pi/2}=\frac1 2\ketbra{1}+\frac 1 2 \cdot \frac{\openone}{2}.\label{eq:examplestate3}
\end{equation}
Numerics show that
\begin{equation}
D_{\rm DPT, sep}(\varrho,\sigma_{\phi})^2=D_{\rm DPT}(\varrho,\sigma_{\phi})^2
\end{equation}
for
\begin{equation}
\phi \ge \phi_0 \approx 0.2946 \pi,\label{eq:phi0}
\end{equation}
while for smaller $\phi$ an entangled $\varrho_{12}$ is cheaper than a separable one.

We can relate $D_{\rm GMPC,sep}(\varrho,\sigma)^2$ to the quantum Fisher information.

\DEFOBS{obs:GMPC_sep}For the modified GMPC distance defined in \EQ{eq:GMPC_distance_sep} the inequality
\begin{equation}\label{eq:GMPC_distance_sep_ineq}
D_{\rm GMPC,sep}(\varrho,\sigma)^2\ge  \frac1 8 \sum_{n=1}^N {\mathcal F}_Q[\varrho,\Hamiltonian_n] +\frac1 8 \sum_{n=1}^N {\mathcal F}_Q[\sigma,\Hamiltonian_n]
\end{equation}
holds, and for $\varrho=\sigma$ and for $N=1$ we have equality in \EQ{eq:GMPC_distance_sep_ineq}.

 \PROOF We can rewrite the optimization problem in \EQ{eq:GMPC_distance_sep} for separable states as
\begin{align}
&D_{\rm GMPC,sep}(\varrho,\sigma)^2\nonumber\\
&\quad=\frac 1 2 \min_{\{p_k,\ket{\Psi_k},\ket{\Phi_k}\}} \sum_{n=1}^N  \sum_k p_k \bigg[\va{\Hamiltonian_n}_{\Psi_k} +  \va{\Hamiltonian_n}_{\Phi_k}\nonumber\\
&\quad\quad\quad\quad+\left(\ex{\Hamiltonian_n}_{\Psi_k}-\ex{\Hamiltonian_n}_{\Phi_k}\right)^2 \bigg]\nonumber\\
&\quad\ge \frac 1 2 \bigg[  \min_{\{p_k,\ket{\Psi_k}\}} \sum_{n=1}^N \sum_k p_k \va{\Hamiltonian_n}_{\Psi_k} \nonumber\\
&\quad\quad\quad\quad+\min_{\{q_k,\ket{\Phi_k}\}} \sum_{n=1}^N \sum_k q_k \va{\Hamiltonian_n}_{\Phi_k}\bigg]\nonumber\\
&\quad\ge \frac 1 2 \bigg[  \sum_{n=1}^N \min_{\{p_k,\ket{\Psi_k}\}}  \sum_k p_k \va{\Hamiltonian_n}_{\Psi_k} \nonumber\\
&\quad\quad\quad\quad+\sum_{n=1}^N\min_{\{q_k,\ket{\Phi_k}\}} \sum_k q_k \va{\Hamiltonian_n}_{\Phi_k}\bigg],
\label{eq:GMPC_distance_sep2}
\end{align}
where for the first optimization the decomposition of $\varrho_{12}$ is given as \EQ{eq:sep_states}, and we have the conditions
\begin{equation}
{\rm Tr}_2(\varrho_{12})=\varrho, \quad {\rm Tr}_1(\varrho_{12})=\sigma. \label{eq:cond_rho_sigma}
\end{equation}
For the second optimization, the condition with $\varrho$ is given in \EQ{eq:purestatedecomp}. For the third optimization, the condition is
\begin{equation}
\sigma=\sum_k q_k \ketbra{\Phi_k},  \label{eq:cond_sigma}
\end{equation}
where for the probabilities $q_k\ge0$ and $\sum_k q_k=1$ hold.
The fourth and fifth optimization are similar to the second and the third one, however, the order of the sum and the minimization is exchanged. From \EQ{eq:GMPC_distance_sep2}, the statement follows using the formula giving the quantum Fisher information with the convex roof of the variance in \EQ{eq:deffqroof}. $\qed$

In \EQ{eq:GMPC_distance_sep2}, in the second and third lines we can see that, when computing $D_{\rm GMPC,sep}(\varrho,\sigma)^2,$ the quantity to be minimized is a weighted sum containing the variances of $\Hamiltonian_n$ for $\ket{\Psi_k}$ and $\ket{\Phi_k},$ and the expression $\left(\ex{\Hamiltonian_n}_{\Psi_k}-\ex{\Hamiltonian_n}_{\Phi_k}\right)^2.$ The two variances, $\va{\Hamiltonian_n}_{\Psi_k}$ and $\va{\Hamiltonian_n}_{\Phi_k},$  are zero if $\ket{\Psi_k}$ and  $\ket{\Phi_k}$ are the eigenstates of $H_n.$ The term $\left(\ex{\Hamiltonian_n}_{\Psi_k}-\ex{\Hamiltonian_n}_{\Phi_k}\right)^2$ is zero if
\begin{equation}
\ex{\Hamiltonian_n}_{\Psi_k}=\ex{\Hamiltonian_n}_{\Phi_k}\label{eq:Ham}
\end{equation}
holds.

Based on \EXAMPLE{ex:ex1} and the proof of \OBS{obs:GMPC_sep}, we find that the GMPC distance can be obtained with an optimization over separable decompositions as
\begin{align}\label{eq:GMPC_distance_sep_recursive}
D_{\rm GMPC,sep}(\varrho,\sigma)^2\quad\quad\quad\nonumber\\=
\min_{\{p_k,\ket{\Psi_k},\ket{\Phi_k}\}}\;&
\sum_k p_k D_{\rm GMPC,sep}(\ket{\Psi_k},\ket{\Phi_k})^2,\nonumber\\
\end{align}
where for the optimization the decomposition of $\varrho_{12}$ is given as \EQ{eq:sep_states}, and we have the conditions on the marginals given in \EQ{eq:cond_rho_sigma}.
This way, we computed the distance for mixed states using the a formula for the distance for pure states and an optimization.  An analogous expressions holds also for $D_{\rm DPT,sep}(\varrho,\sigma)^2.$

Let us see now the consequences of the above observations for the self-distance, which, unlike in the classical case, can be nonzero. It is obtained as
\begin{align}
 D_{\rm GMPC, sep}(\varrho,\varrho)^2 &=& \min_{\{p_k,\ket{\Psi_k}\}}\;\sum_{n=1}^N \sum_k p_k \va{\Hamiltonian_n}_{\Psi_k},\nonumber\\
 \label{eq:DGMPC_sep_rhorho}
\end{align}
and the condition for the minimization is given in \EQ{eq:purestatedecomp}.  Then, based on \EQ{eq:DGMPC2DGMPCsep2}, it follows that the self-distance for the GMPC distance is bounded from above as
\begin{align}
 D_{\rm GMPC}(\varrho,\varrho)^2 &\le& \min_{\{p_k,\ket{\Psi_k}\}}\;\sum_{n=1}^N \sum_k p_k \va{\Hamiltonian_n}_{\Psi_k}.\nonumber\\
 \label{eq:DGMPC_sep_rhorho2}
\end{align}
Let us examine some properties of $D_{\rm GMPC, sep}(\varrho,\varrho)^2.$ From \EQ{eq:DGMPC_sep_rhorho} it follows that for the self-distance for $N=1$ for a given $H_1$
\begin{equation}
D_{\rm GMPC, sep}(\varrho,\varrho)^2=0
\end{equation}
holds, if and only if
\begin{equation}
[\varrho,H_1]=0.\label{eq:commutators}
\end{equation}

After studying the self-distance, let us look now for similar relations for the distance between two different states.  We find that for $N=1$ for a given $H_1$
\begin{equation}
D_{\rm GMPC, sep}(\varrho,\sigma)^2=0
\end{equation}
holds, if and only if
\begin{equation}
[\varrho,H_1]=[\sigma,H_1]=0\label{eq:commutators2}
\end{equation}
is satisfied and if there is $\{p_k,\ket{\Psi_k},\ket{\Phi_k}\}$ corresponding to the optimum such that \EQ{eq:Ham} holds for all $k.$  In this case, if $H_1$ has a non-degenerate spectrum then $\ket{\Psi_k}$ and $\ket{\Phi_k}$ are the eigenvectors of $\varrho, \sigma$ and $H_1,$  and  thus even $[\varrho,\sigma]=0$ holds. Finally, based on \OBS{obs:distance_sep_equal}, analogous relations hold for $D_{\rm DPT, sep}(\varrho,\sigma)^2.$

Let us look for a relation between the distance and the self-distance.  From the first inequality in \EQ{eq:GMPC_distance_sep2}, it follows that for any $\varrho$ and $\sigma$
\begin{align}
&D_{\rm GMPC, sep}(\varrho,\sigma)^2 \ge\nonumber\\
&\quad\quad\quad \frac1 2 \left[  D_{\rm GMPC, sep}(\varrho,\varrho)^2 + D_{\rm  GMPC, sep}(\sigma,\sigma)^2 \right]\label{eq:DGMPC}\nonumber\\
\end{align}
holds.

For a local dimension $d>2,$ the optimization over separable states is difficult to carry out numerically. Thus, it is reasonable to define $D_{\rm DPT, PPT}(\varrho,\sigma)^2$ and $D_{\rm GMPC, PPT}(\varrho,\sigma)^2$ that need an optimization over PPT states rather than separable states. It is possible to prove that the two new quantities are equal to each other.

\DEFOBS{obs:distance_ppt_equal}The two quantum Wasserstein distance measures are equal to each other
\begin{equation}\label{eq:distance_ppt_ineq}
D_{\rm DPT, PPT}(\varrho,\sigma)^2=D_{\rm GMPC, PPT}(\varrho,\sigma)^2.
\end{equation}

 \PROOF We have to follow ideas similar to the ones in \OBS{obs:distance_sep_equal}. In particular, we need to use that $\varrho_{12}$ is a PPT quantum state if and only if
$\varrho_{12}^{T_1}$ is a PPT quantum state.$ \qed$

In order to define further Wasserstein distance measures based on an optimization over other supersets of separable states, we need to know the separability criterion based on symmetric extensions \cite{Doherty2002Distinguishing, Doherty2004Complete,Doherty2005Detecting}. A given bipartite state $\varrho_{AB}$ is said to have a $n:m$ symmetric extension if it can be written as the reduced state of a multipartite state  $\varrho_{A_1..A_nB_1..B_m},$ which is symmetric under $A_k\leftrightarrow A_l$ for all $k\ne l$ and under $B_{k'}\leftrightarrow B_{l'}$ for all $k'\ne l'.$ If we also require that the state is PPT for all bipartitions, then the state has a PPT symmetric extension. The requirement of having a PPT symmetric extension for $n=1$ and $m=1$ is equivalent to the PPT condition, while for $n>1$ or $m>1$ the condition is stronger. Bipartite separable states have such extensions for arbitrarily large $n$ and $m,$ while the lack of such  an extension for some $m$ and $n$ signals the presence of entanglement. In particular, a state is separable if and only if there is an extension for any $n$ and for $m=1.$

Let us define $D_{{\rm DPT, PPT}n}(\varrho,\sigma)^2$ and $D_{{\rm GMPC, PPT}n}(\varrho,\sigma)^2$ based on an optimization over quantum states with a PPT symmetric extension for given $n$ and for $m=1.$
\begin{equation}
D_{{\rm DPT, PPT}n}(\varrho,\sigma)^2\le D_{{\rm DPT, PPT}n'}(\varrho,\sigma)^2\label{eq:PPTnPPTnprime}
\end{equation}
if $n'>n.$ Based on \OBSS{obs:distance_sep_equal} and \ref{obs:distance_ppt_equal}, we can also see that
\begin{equation}
D_{{\rm DPT, PPT}n}(\varrho,\sigma)^2=D_{{\rm GMPC, PPT}n}(\varrho,\sigma)^2\label{eq:PPTnGMPCn}
\end{equation}
for every $n.$

\section{Variance-like quantities}
\label{sec:var}

When we compare the expression in \EQ{eq:deffqroof3} defining the quantum Fisher information and the other expression in \EQ{eq:VAR_max_sep_states} defining the variance, we can see that the main difference is that the minimization is replaced by a maximization. We also showed a similar relation between the entanglement of formation given in \EQ{eq:EF}, and the entanglement of assistance given in \EQ{eq:EA}. Based on this observation, we can define variance-like quantities from the various forms of quantum Wasserstein distance by replacing the minimization by maximization.

Such a variance-like quantity can be interpreted in the framework of transport problems as follows. The quantum Wasserstein distance determines the smallest cost possible for the transport problem by a minimization. The variance-like quantities presented in this section determine the largest cost possible for the transport problem. Knowing the largest possible cost is useful when judging how close the cost of a given transport plan is to the optimal cost.

Let us now define the first variance-like quantity.

\DEFDEFINITION{def:V} From the  GMPC distance with an optimization restricted over separable states given in \EQ{eq:GMPC_distance_sep}, we obtain the following variance like quantity
\begin{align}
V_{\rm GMPC,sep} (\varrho,\sigma)\quad\quad\quad
\nonumber\\
=\frac 1 2
\max_{ \varrho_{12}}\sum_{n=1}^N\;&
\trace[(\Hamiltonian_n\otimes \openone-\openone\otimes \Hamiltonian_n)^2  \varrho_{12} ],\nonumber\\
\textrm{s.~t. }&
\varrho_{12}\in\mathcal S,\label{eq:GMPC_variance_sep}  \nonumber\\
& {\rm Tr}_2(\varrho_{12})=\varrho, \nonumber\\
& {\rm Tr}_1(\varrho_{12})=\sigma.
\end{align}
Clearly, the inequality
\begin{equation}
V_{\rm GMPC,sep} (\varrho,\sigma)\ge D_{\rm GMPC,sep}(\varrho,\sigma)^2\label{eq:VGMPCsepDGMPCsep2}
\end{equation}
holds, since since on the left-hand side of \EQ{eq:VGMPCsepDGMPCsep2} we maximize over a set of quantum states while on the right-hand side we minimize over the same set.

We can define analogously
$V_{\rm DPT, sep} (\varrho,\sigma),$
$V_{\rm GMPC} (\varrho,\sigma),$ and
$V_{\rm DPT}(\varrho,\sigma),$
modifying the definition of $D_{\rm DPT, sep}(\varrho,\sigma)^2, $ $D_{\rm GMPC}(\varrho,\sigma)^2,$ and $D_{\rm DPT}(\varrho,\sigma)^2,$ respectively. In all these cases, a relation analogous to the one in \EQ{eq:VGMPCsepDGMPCsep2} can be obtained.

The value of $V_{\rm GMPC,sep} (\varrho,\sigma)$ is related to the variance.

\DEFOBS{obs:GMPC_distance_bound}The GMPC variance defined in \EQ{eq:GMPC_variance_sep} is bounded from below as
\begin{equation}\label{eq:GMPC_variance_sep_ineq}
V_{\rm GMPC,sep}(\varrho,\sigma) \ge \frac1 2 \sum_{n=1}^N  \va{\Hamiltonian_n}_{\varrho}+\va{\Hamiltonian_n}_{\sigma}.
\end{equation}

 \PROOF We can rewrite the expression to be computed for $V_{\rm GMPC,sep}(\varrho,\sigma)$  as
\begin{align}
&\frac1 2\max_{\varrho_{12}\in\mathcal S}\;\sum_{n=1}^N\trace[(\Hamiltonian_n\otimes \openone-\openone\otimes \Hamiltonian_n)^2  \varrho_{12} ]\nonumber\\
&\quad\quad=\frac1 2\sum_{n=1}^N\va{H_n}_{\varrho}+\va{H_n}_{\sigma}+(\ex{\Hamiltonian_n}_{\varrho}-\ex{\Hamiltonian_n}_{\sigma})^2\nonumber\\
&\quad\quad-\min_{\varrho_{12}\in\mathcal S}\;C_{\varrho_{12}},\label{eq:tobeoptimzied}
\end{align}
where $\varrho_{12}$ has marginals $\varrho$ and $\sigma,$ as given in  \EQ{eq:cond_rho_sigma}, and we define
\begin{equation}
C_{\varrho_{12}}=\sum_{n=1}^N \ex{H_n\otimes H_n}_{\varrho_{12}}-\ex{H_n}_{\varrho}\ex{H_n}_{\sigma}.\label{eq:C}
\end{equation}
Since the product state $\varrho\otimes\sigma$ is separable and fulfills the conditions on the marginals, it is clear that
\begin{equation}\label{eq:ineq}
\min_{ \varrho_{12}\in\mathcal{S}}C_{\varrho_{12}}\le C_{\varrho\otimes\sigma} =0.
\end{equation}
Substituting \EQ{eq:ineq} into \EQ{eq:tobeoptimzied}, using the fact $(\ex{\Hamiltonian_n}_{\varrho}-\ex{\Hamiltonian_n}_{\sigma})^2\ge0,$ we can prove the observation. $\qed$

We can even obtain an upper bound.

\DEFOBS{obs:GMPC_distance_upper_bound}The GMPC variance defined in \EQ{eq:GMPC_variance_sep} is bounded from above as
\begin{align}\label{eq:GMPC_variance_sep_ineq_upperbound}
V_{\rm GMPC,sep}(\varrho,\sigma) &\le  \sum_{n=1}^N  \va{\Hamiltonian_n}_{\varrho}+\va{\Hamiltonian_n}_{\sigma}\nonumber\\
&+\ex{H_n}_{\varrho}^2+\ex{H_n}_{\sigma}^2.
\end{align}

 \PROOF We can find an upper bound for the expression to be computed for $V_{\rm GMPC,sep}(\varrho,\sigma)$  as
\begin{align}
&\frac1 2\max_{\varrho_{12}\in\mathcal S}\;\sum_{n=1}^N\trace[(\Hamiltonian_n\otimes \openone-\openone\otimes \Hamiltonian_n)^2  \varrho_{12} ]\nonumber\\
&\quad\quad\quad\le\sum_{n=1}^N\ex{\Hamiltonian_n^2}_{\varrho}+\ex{\Hamiltonian_n^2}_{\sigma},
\label{eq:tobeoptimzied_upperbound}
\end{align}
where $\varrho_{12}$ has marginals $\varrho$ and $\sigma,$ as given in  \EQ{eq:cond_rho_sigma}, and we used that
\begin{equation}
\Hamiltonian_n^2\otimes \openone+\openone\otimes \Hamiltonian_n^2\ge 2\Hamiltonian_n \otimes \Hamiltonian_n.
\end{equation}
$\qed$

Let us see some concrete examples.

\DEFEXAMPLE{ex:ex5}Interestingly, the quantity
\begin{equation}
\mathcal V:=V_{\rm GMPC, sep}(\varrho,\sigma)
\end{equation}
can be larger or smaller than or equal to
\begin{equation}
\overline{\mathcal V}:=\frac1 2 \left[  V_{\rm GMPC, sep}(\varrho,\varrho) + V_{\rm  GMPC, sep}(\sigma,\sigma) \right],
\end{equation}
c. f. \EQ{eq:DGMPC}. We consider $N=1$ and $H_1=\sigma_z.$
The three possibilities above are realized by the following states. For the state
\begin{align}
\varrho&=\ketbra{0},\nonumber\\
\sigma&={\rm diag}(0.25,0.75),
\end{align}
we have $\mathcal V>\overline{\mathcal V}.$ For the following state
\begin{align}
\varrho&=\ketbra{0},\nonumber\\
\sigma&={\rm diag}(0.75,0.25),
\end{align}
we have $\mathcal V=\overline{\mathcal V}.$ Finally, for the following state
\begin{align}
\varrho&=\openone/2,\nonumber\\
\sigma&=\left(\begin{array}{cc}0.75 & 0.40 \\0.40 & 0.25\end{array}\right),
\end{align}
we have $\mathcal V<\overline{\mathcal V}.$ Let us look now for larger systems. For systems with a local dimension $d=3$ and for $H_1={\rm diag}(-1,0,1),$ $N=1$ and for
\begin{align}
\varrho&=\openone/3,\nonumber\\
\sigma&={\rm diag}(1,0,0),\label{eq:d3example}
\end{align}
we have $\mathcal V<\overline{\mathcal V}.$ Note that in all examples where $\varrho, \sigma$ and $H_1$ were all diagonal, there is an optimal diagonal $\varrho_{12},$ essentially corresponding to the classical case.

The two different variance-like quantities, defined with the transpose and without it, respectively, are equal to each other.

\DEFOBS{obs:VAR_sep_equal}The two types of quantum Wasserstein variance are equal to each other
\begin{equation}
V_{\rm DPT, sep}(\varrho,\sigma) = V_{\rm GMPC, sep}(\varrho,\sigma).
\end{equation}
 \PROOF The proof is analogous to that of \OBS{obs:distance_sep_equal}. $\qed$

Let us calculate $V_{\rm DPT, sep}(\varrho,\sigma)$ for some concrete examples.

\DEFEXAMPLE{ex:ex6}Let us consider the case when $\varrho=\ketbra{\Psi}$ is a pure state of any dimension and $\sigma$ is an arbitrary density matrix of the same dimension. Then, when computing the various types of quantum Wasserstein distance and quantum Wasserstein variance between $\varrho$ and $\sigma,$ the state $\varrho_{12}$ in the optimization is constrained to be the tensor product of the two density matrices. Hence, it follows that
\begin{equation}
V_{\rm DPT}(\ketbra{\Psi},\sigma)=D_{\rm DPT}(\ketbra{\Psi},\sigma)^2
\label{eq:VHn}
\end{equation}
holds, and analogous equations hold for the quantities $V_{\rm DPT, sep}(\ketbra{\Psi},\sigma), V_{\rm GMPC, sep}(\ketbra{\Psi},\sigma),$ and $V_{\rm GMPC}(\ketbra{\Psi},\sigma),$ where the Wasserstein distance measures for this case are given in \EQ{eq:D2_pure_mixed}.

\DEFEXAMPLE{ex:ex7}For $\varrho=\sigma=\ketbra{\Psi},$ and for $N=1$ for a given $H_1$ we obtain
\begin{equation}
V_{\rm DPT}(\ketbra{\Psi},\ketbra{\Psi})=\va{H_1}_{\Psi}.
\end{equation}

\DEFEXAMPLE{ex:ex8}Let us consider the single-qubit states given in \EQ{eq:totallymixedrhosigma}. Let us take $N=1$ and $\Hamiltonian_1=\sigma_z.$ Then, when computing $V_{\rm GMPC,sep}(\varrho,\sigma),$ the optimum is attained by the separable state
\begin{equation}
\varrho_{12}=\frac1 2\ketbra{01} + \frac1 2 \ketbra{10}, \label{eq:rho12_optim_var}
\end{equation}
[c.~f. \EQ{eq:rho12_optim}] and we have
\begin{equation}
V_{\rm GMPC,sep}\left(\frac1 2 \openone,\frac1 2 \openone\right)=2.
\end{equation}
(See also \EXAMPLE{ex:ex0}.) If $\Hamiltonian_1$ is a different operator then the state $\varrho_{12}$ corresponding to the maximum will be different. It can be obtained from the state given in \EQ{eq:rho12_optim_var} with local unitaries. For instance, for $\Hamiltonian_1=\sigma_x,$ the optimum is reached by the separable state
\begin{equation}
\varrho_{12}=\frac1 2 \ketbra{01}_x + \frac1 2 \ketbra{10}_x, \label{eq:rho12_optim_x_var}
\end{equation}
where $\ket{.}_x$ is a state given in the $x$-basis.

\section{Quantum Wasserstein distance and entanglement criteria}
\label{sec:wassent}

\EXAMPLE{ex:ex4} highlighted that in certain cases the minimum for separable states is larger than the minimum for general states. In this case, entanglement can help to decrease the quantum Wasserstein distance. In this section, we analyze the relation of the quantum Wasserstein distance and entanglement conditions on the optimal $\varrho_{12}$ couplings.

First, we make the following simple observation.

\DEFOBS{obs:entanglement_detetcion}If
\begin{equation}
D_{\rm DPT, sep}(\varrho,\sigma)^2>D_{\rm DPT}(\varrho,\sigma)^2\label{eq:Dineq1}
\end{equation}
holds then all $\varrho_{12}$ states that minimize the cost  for a given $\varrho$ and $\sigma,$ when computing $D_{\rm DPT}(\varrho,\sigma)^2,$  are entangled. In short, all optimal $\varrho_{12}$ states are entangled. The situation is analogous if
\begin{equation}
D_{\rm GMPC, sep}(\varrho,\sigma)^2>D_{\rm GMPC}(\varrho,\sigma)^2.\label{eq:Dineq2}
\end{equation}
Then, all optimal $\varrho_{12}$ states for a given $\varrho$ and $\sigma,$ when computing $D_{\rm GMPC}(\varrho,\sigma)^2,$ are entangled. Thus, we can even use the quantum Wasserstein distance as an entanglement criterion detecting entanglement in the optimal $\varrho_{12}$ states.

We can even consider conditions with the self-distance. Based on \EQS{eq:DrhoI}, \eqref{eq:self_distance_GMPC}, and \eqref{eq:distance_sep_ineq}, we find that for the $N=1$ case
\begin{equation}
D_{\rm DPT, sep}(\varrho,\varrho)^2>D_{\rm DPT}(\varrho,\varrho)^2\label{eq:Dineq1b}
\end{equation}
is equivalent to a relation between the quantum Fisher information and the Wigner-Yanase skew information
\begin{equation}
\frac{F_Q[\varrho,H_1]}4 > I_\varrho(H_1).\label{eq:FQIrho}
\end{equation}
Thus, the condition in \EQ{eq:FQIrho} implies that the optimal $\varrho_{12}$ states, obtained when computing $D_{\rm DPT}(\varrho,\varrho)^2,$ are all entangled.

Let us now use entanglement criteria to construct relations for the quantum Wasserstein distance, that can verify that the coupling $\varrho_{12}$ is entangled. If the inequality given in \EQ{eq:entond} holds for separable states, so does the inequality
\begin{equation}
\sum_{n=1}^{d^2-1} \ex{(G_n^T \otimes \openone - \openone \otimes G_n)^2} \ge 4(d-1),\label{eq:entond2}
\end{equation}
since the left-hand side of \EQ{eq:entond2} is never smaller than the left-hand side of \EQ{eq:entond}. Any state that violates the inequality in \EQ{eq:entond2} is entangled. It can be shown that \EQ{eq:entond2} is a tight inequality for separable states as follows. Based on \EQ{eq:entond1}, we see that for pure product states of the form
\begin{equation}
(\ketbra{\Psi})^T\otimes\ketbra{\Psi},
\end{equation}
for the second moments
\begin{equation}
\sum_{n=1}^{d^2-1} \ex{(G_n^T \otimes \openone - \openone \otimes G_n)^2} = 4(d-1)\label{eq:entond2222}
\end{equation}
holds since for this state
\begin{equation}
\ex{G_n^T \otimes \openone - \openone \otimes G_n} = 0
\end{equation}
for all $n.$

Next, we will define a quantum Wasserstein distance related to the entanglement condition in \EQ{eq:entond2}.

\DEFOBS{obs:wasserstein_witness}Let us consider $d$-dimensional systems with
\begin{equation}
\Hamiltonian_n=G_n
\end{equation}
for $n=1,2,...,d^2-1.$ If
\begin{equation}
D_{\rm DPT}^{\{G_1,G_2,...,G_{d^2-1}\}}(\varrho,\sigma)^2<2(d-1)
\end{equation}
holds, then all optimal $\varrho_{12}$ states are entangled. Here, for clarity, we give explicitly the observables used to define the distance in the superscript.

Clearly, since when calculating $D_{\rm DPT, sep}(\varrho,\sigma)^2,$ we optimize over separable states, we have
\begin{equation}
D_{\rm DPT, sep}^{\{G_1,G_2,...,G_{d^2-1}\}}(\varrho,\sigma)^2\ge2(d-1).
\end{equation}
Thus, independently from what $\varrho$ and $\sigma$ are, their distance $D_{\rm DPT, sep}^{\{G_1,G_2,...,G_{d^2-1}\}}(\varrho,\sigma)^2$ cannot be smaller than a bound. This is true even if $\varrho=\sigma.$

Let us now use another entanglement condition to construct relations for the quantum Wasserstein distance that can verify that the coupling is entangled. We know that the inequality given in \EQ{eq:2qubitb} holds for separable states and it is tight. Based on these, we can obtain the following bounds on the quantum Wasserstein distance.

\DEFOBS{obs:wasserstein_witness_JxJyJz}Let us choose the set of operators as
\begin{equation}
\{\Hamiltonian_n\}=\{j_x,j_y,j_z\}.
\end{equation}
Then, if the inequality
\begin{equation}
D_{\rm DPT}^{\{j_x,j_y,j_z\}}(\varrho,\sigma)^2<j
\end{equation}
holds, then all optimal $\varrho_{12}$ states are entangled.

Clearly, since when calculating $D_{\rm DPT, sep}(\varrho,\sigma)^2,$ we optimize over separable states, we have
\begin{equation}
D^{\{j_x,j_y,j_z\}}_{\rm DPT, sep}(\varrho,\sigma)^2\ge j.
\end{equation}
Thus, again, independently from what $\varrho$ and $\sigma$ are, their distance $D^{\{j_x,j_y,j_z\}}_{\rm DPT, sep}(\varrho,\sigma)^2$ cannot be smaller than a bound. This is true even if $\varrho=\sigma.$

So far we studied the relation of the quantum Wasserstein distance to entanglement. Next, let us consider the relation of $V_{\rm DPT}(\varrho,\sigma)$ and $V_{\rm GMPC}(\varrho,\sigma)$ to entanglement.

\DEFOBS{obs:wasserstein_var_ent}If
\begin{equation}
V_{\rm DPT, sep}(\varrho,\sigma)<V_{\rm DPT}(\varrho,\sigma)
\end{equation}
holds then all optimal $\varrho_{12}$ states for a given $\varrho$ and $\sigma,$ when computing $V_{\rm DPT}(\varrho,\sigma),$ are entangled. The situation is analogous if
\begin{equation}
V_{\rm GMPC, sep}(\varrho,\sigma)<V_{\rm GMPC}(\varrho,\sigma).
\end{equation}

Next, we will determine a set of $\Hamiltonian_n$ operators that can be used efficiently to detect entanglement with the quantum Wasserstein variance. For that, we need to know that for separable states the inequality given in \EQ{eq:ineqsep} holds.

\DEFOBS{obs:wasserstein_var_ent_cond}Let us consider an example with $d=2$ and $\{H_n\}=\{\sigma_x,\sigma_y\}.$ If the condition
\begin{equation}
V_{\rm GMPC}^{\{\sigma_x,\sigma_y\}}(\varrho,\sigma)>3
\end{equation}
holds, then all optimal $\varrho_{12}$ states are entangled. Clearly, since when calculating $V_{\rm DPT, sep}(\varrho,\sigma),$ we optimize over separable states, we have
\begin{equation}
V_{\rm GMPC, sep}^{\{\sigma_x,\sigma_y\}}(\varrho,\sigma)\le 3.
\end{equation}

Let us see now a complementary relation for $D_{\rm GMPC}(\varrho,\sigma)^2$ and $D_{\rm GMPC,sep}(\varrho,\sigma)^2.$ They will use the same $H_n$ operators that appear in \OBS{obs:wasserstein_var_ent_cond}. We need to know that for separable states the inequality given in \EQ{eq:2var2} holds.

\DEFOBS{obs:wasserstein_dist_ent2}Let us consider $d=2$ and $\{H_n\}=\{\sigma_x,\sigma_y\}.$ If
\begin{equation}
D_{\rm GMPC}^{\{\sigma_x,\sigma_y\}}(\varrho,\sigma)^2<1
\end{equation}
then all optimal $\varrho_{12}$ states are entangled. Clearly, since when calculating $D_{\rm DPT, sep}(\varrho,\sigma),$ we optimize over separable states, we have
\begin{equation}
D_{\rm GMPC, sep}^{\{\sigma_x,\sigma_y\}}(\varrho,\sigma)^2\ge 1.
\end{equation}

Statements analogous to those of \OBSS{obs:wasserstein_var_ent_cond} and \ref{obs:wasserstein_dist_ent2} can be formulated, with identical bounds, for $V_{\rm DPT}(\varrho,\sigma)^2,$ $V_{\rm DPT,sep}(\varrho,\sigma)^2,$ $D_{\rm DPT}(\varrho,\sigma)^2,$ and $D_{\rm DPT,sep}(\varrho,\sigma)^2.$

\section{Optimization of the variance over the two-copy space}

\label{sec:variance_instead_of_second_moment}

In this section, we examine the quantity that we obtain after replacing the second moment by a variance in the optimization in the definition of $D_{\rm GMPC,sep}(\varrho,\sigma)^2$ given in \EQ{eq:GMPC_distance_sep} and in the definition of $V_{\rm GMPC,sep}(\varrho,\sigma)$ in \EQ{eq:GMPC_variance_sep}. Analogous ideas work also for the other types of quantum Wasserstein distance and quantum Wasserstein variance defined before. We will show that such quantities have interesting properties.

\DEFDEFINITION{def:tildeD_GMC_sep}After replacing the second moment by a variance in the optimization in the definition of $D_{\rm GMPC,sep}(\varrho,\sigma)^2$ given in \EQ{eq:GMPC_distance_sep}, we define
\begin{align}
\tilde D_{\rm GMPC,sep}(\varrho,\sigma)^2\quad\quad\quad\nonumber\\=\frac 1 2
\min_{ \varrho_{12}}\;\sum_{n=1}^N\;&
\vasq{(\Hamiltonian_n\otimes \openone-\openone\otimes \Hamiltonian_n)}_{\varrho_{12}},\nonumber\\
\textrm{s.~t. }&
\varrho_{12}\in\mathcal S,\label{eq:tilde_GMPC_distance_sep}  \nonumber\\
& {\rm Tr}_2(\varrho_{12})=\varrho, \nonumber\\
& {\rm Tr}_1(\varrho_{12})=\sigma.
\end{align}

Let us see some properties of the quantity we have just introduced. In general, for mixed states,
\begin{align}
\tilde D_{\rm GMPC,sep}(\varrho,\sigma)^2&=D_{\rm GMPC,sep}(\varrho,\sigma)^2\nonumber\\
&-\frac1 2\sum_n (\ex{H_n}_{\varrho}-\ex{H_n}_{\sigma})^2\label{eq:tiltdeDGMPCSep_ineq}\nonumber\\
\end{align}
holds, hence clearly
\begin{equation}
\tilde D_{\rm GMPC,sep}(\varrho,\sigma)^2\le D_{\rm GMPC,sep}(\varrho,\sigma)^2.
\end{equation}
Due to the relation in \EQ{eq:tiltdeDGMPCSep_ineq}, for the self-distance we have
\begin{equation}
\tilde D_{\rm GMPC,sep}(\varrho,\varrho)^2=D_{\rm GMPC,sep}(\varrho,\varrho)^2,
\end{equation}
where $D_{\rm GMPC,sep}(\varrho,\sigma)^2$ is given in \EQ{eq:GMPC_distance_sep}.

We can write the expression to be optimized as
\begin{align}
&\sum_{n=1}^N\vasq{(\Hamiltonian_n\otimes \openone-\openone\otimes \Hamiltonian_n)}_{\varrho_{12}}\nonumber\\
&\quad\quad\quad=\sum_{n=1}^N[\va{H_n}_{\varrho}+\va{H_n}_{\sigma}]-2C_{\varrho_{12}},
\end{align}
where $\varrho_{12}$ has marginals $\varrho$ and $\sigma,$ as given in  \EQ{eq:cond_rho_sigma}, and we define $C_{\varrho_{12}}$ as in \EQ{eq:C}. For pure states, $\varrho=\ketbra{\Psi}$ and  $\sigma=\ketbra{\Phi}$, we have $C=0,$ and hence
\begin{align}
&\tilde D_{\rm GMPC,sep}(\ketbra{\Psi},\ketbra{\Phi})^2\nonumber\\&\quad\quad\quad\quad=\frac1 2\bigg[\tilde D_{\rm GMPC,sep}(\ketbra{\Psi},\ketbra{\Psi})^2\nonumber\\
&\quad\quad\quad\quad\quad+\tilde D_{\rm GMPC,sep}(\ketbra{\Phi},\ketbra{\Phi})^2\bigg].\label{eq:tildeDGMPCsep_sum}
\end{align}

We can present lower bounds on the distance.

\DEFOBS{obs:tildeGMPC_sep}The modified GMPC distance defined in \EQ{eq:tilde_GMPC_distance_sep} is bounded from below as
\begin{equation}\label{eq:tilde_GMPC_distance_sep_ineq}
\tilde D_{\rm GMPC,sep}(\varrho,\sigma)^2\ge  \frac1 8 \sum_{n=1}^N {\mathcal F}_Q[\varrho,\Hamiltonian_n] +\frac1 8 \sum_{n=1}^N {\mathcal F}_Q[\sigma,\Hamiltonian_n],
\end{equation}
while for $\varrho=\sigma$ we have equality for $N=1$ in \EQ{eq:tilde_GMPC_distance_sep_ineq}.

 \PROOF We can rewrite the optimization problem in \EQ{eq:tilde_GMPC_distance_sep} based on \EQS{eq:GMPC_distance_sep2} and \eqref{eq:tiltdeDGMPCSep_ineq} as
\begin{align}
&\tilde D_{\rm GMPC,sep}(\varrho,\sigma)^2\nonumber\\
&\quad=\frac 1 2\min_{\{p_k,\ket{\Psi_k},\ket{\Phi_k}\}}\;\sum_{n=1}^N  \sum_k p_k \bigg[\va{\Hamiltonian_n}_{\Psi_k} +  \va{\Hamiltonian_n}_{\Phi_k}\nonumber\\
&\quad\quad\quad\quad+\left(\ex{\Hamiltonian_n}_{\Psi_k}-\ex{\Hamiltonian_n}_{\Phi_k}\right)^2 - \left(\ex{\Hamiltonian_n}_{\varrho}-\ex{\Hamiltonian_n}_{\sigma}\right)^2 \bigg]\nonumber\\
&\quad\ge \frac 1 2 \bigg[ \min_{\{p_k,\ket{\Psi_k}\}}\;\sum_{n=1}^N \sum_k p_k \va{\Hamiltonian_n}_{\Psi_k} \nonumber\\
&\quad\quad\quad\quad+\min_{\{q_k,\ket{\Phi_k}\}}\;\sum_{n=1}^N \sum_k q_k \va{\Hamiltonian_n}_{\Phi_k}\bigg]\nonumber\\
&\quad\ge \frac 1 2 \bigg[ \sum_{n=1}^N\min_{\{p_k,\ket{\Psi_k}\}}\;\sum_k p_k \va{\Hamiltonian_n}_{\Psi_k}\nonumber\\
&\quad\quad\quad\quad+\sum_{n=1}^N\min_{\{q_k,\ket{\Phi_k}\}}\;\sum_k q_k \va{\Hamiltonian_n}_{\Phi_k}\bigg],
\label{eq:tilde_GMPC_distance_sep2}
\end{align}
where we used in the first inequality that for real $x_k$
\begin{equation}
\sum_k p_k x_k^2 \ge \left( \sum_k p_k x_k\right)^2
\end{equation}
holds due to the fact that $f(x)=x^2$ is convex. In \EQ{eq:tilde_GMPC_distance_sep2}, in the first optimization the decomposition of $\varrho_{12}$ is given as \EQ{eq:sep_states}, and we have the conditions given in \EQ{eq:cond_rho_sigma}. For the second optimization, the condition is \EQ{eq:purestatedecomp}. For the third optimization, the condition is given in \EQ{eq:cond_sigma}, where for the probabilities $q_k\ge0$ and $\sum_k q_k=1$ hold. The fourth and fifth optimization are similar to second and the third one, however, the order of the sum and the minimization is exchanged. From \EQ{eq:tilde_GMPC_distance_sep2}, the statement follows using the formula that obtains the quantum Fisher information with a convex roof of the variance given in \EQ{eq:deffqroof}.$\qed$

We can define another variance-like quantity.

\DEFDEFINITION{def:tildeV_GMC_sep}Analogously, we can define the quantity that we obtain after replacing the second moment by a variance in the optimization in the definition of $V_{\rm GMPC,sep}(\varrho,\sigma)^2$ in \EQ{eq:GMPC_variance_sep}, as
\begin{align}
\tilde V_{\rm GMPC,sep} (\varrho,\sigma)\quad\quad\quad\nonumber\\=\frac 1 2
\max_{\varrho_{12}}\sum_{n=1}^N\;&
\vasq{(\Hamiltonian_n\otimes \openone-\openone\otimes \Hamiltonian_n)}_{\varrho_{12}},\nonumber\\
\textrm{s.~t. }&
\varrho_{12}\in\mathcal S,\label{eq:tilde_GMPC_variance_sep}  \nonumber\\
& {\rm Tr}_2(\varrho_{12})=\varrho, \nonumber\\
& {\rm Tr}_1(\varrho_{12})=\sigma.
\end{align}

Let us see now some properties of $\tilde V_{\rm GMPC,sep} (\varrho,\sigma).$ In general,
\begin{align}\label{eq:V-tilde}
\tilde V_{\rm GMPC,sep}(\varrho,\sigma)&=V_{\rm GMPC,sep}(\varrho,\sigma)\nonumber\\
&-\frac{1}{2}\sum_n (\ex{H_n}_{\varrho}-\ex{H_n}_{\sigma})^2
\end{align}
holds, hence clearly
\begin{equation}
\tilde V_{\rm GMPC,sep}(\varrho,\sigma)\le V_{\rm GMPC,sep}(\varrho,\sigma).
\end{equation}
Analogously to \EQ{eq:tildeDGMPCsep_sum}, for pure states
\begin{align}
&\tilde V_{\rm GMPC,sep}(\ketbra{\Psi},\ketbra{\Phi})\nonumber\\&\quad\quad\quad\quad=\frac1 2\bigg[\tilde V_{\rm GMPC,sep}(\ketbra{\Psi},\ketbra{\Psi})\nonumber\\
&\quad\quad\quad\quad\quad+\tilde V_{\rm GMPC,sep}(\ketbra{\Phi},\ketbra{\Phi})\bigg]
\end{align}
holds.

\DEFEXAMPLE{ex:ex9}Let us consider the case when $\varrho=\ketbra{\Psi}$ is a pure state of any dimension and $\sigma$ is an arbitrary density matrix of the same dimension. Then, when computing the various types of quantum Wasserstein distance and quantum Wasserstein variance between $\varrho$ and $\sigma,$ the state $\varrho_{12}$ in the optimization is constrained to be the tensor product of the two density matrices. Hence, based on \EQS{eq:tiltdeDGMPCSep_ineq} and \eqref{eq:V-tilde}, it follows that
\begin{align}
\tilde D_{\rm GMPC, sep}(\ketbra{\Psi},\sigma)&=\nonumber\\
\tilde V_{\rm GMPC, sep}(\ketbra{\Psi},\sigma)&=\frac1 2 \sum_{n=1}^N  \va{\Hamiltonian_n}_{\Psi}+\va{\Hamiltonian_n}_{\sigma}\nonumber\\
\end{align}
holds, c. f. \OBS{obs:GMPC_distance_bound}.

Finally, one can define similarly  $\tilde D_{\rm DPT, sep}(\varrho,\sigma)^2,$ $\tilde D_{\rm GMPC}(\varrho,\sigma)^2,$ $\tilde D_{\rm DPT}(\varrho,\sigma)^2,$ $\tilde V_{\rm DPT, sep}(\varrho,\sigma),$ $\tilde V_{\rm GMPC}(\varrho,\sigma),$ and $\tilde V_{\rm DPT}(\varrho,\sigma).$

\section{Optimization over other subsets of physical states}
\label{sec:other}

So far we considered Wasserstein distance based on an optimization over all bipartite physical states, separable states, PPT states, and states with a PPT symmetric extension considered in \SEC{sec:Wasserstein}. In this section we examine other convex sets of quantum states. We will also discuss some relevant couplings. By optimizing over a convex set different from the ones we have considered, we will obtain a Wasserstein distance with a different self-distance.

Let us consider the set of couplings for which the quantum discord is zero \cite{Ollivier2001Quantum,Henderson2001Classical,Bera2017Quantum}. If we assume that
\begin{align} {\rm Tr}_2(\varrho_{12})&=\varrho, \nonumber\\
 {\rm Tr}_1(\varrho_{12})&=\sigma
\end{align}
hold, then an element of the set is of the form
\begin{equation}
\varrho_{12}=\sum_k \lambda_k \ketbra{k} \otimes \sigma_k,\label{eq:rho12cc}
\end{equation}
where the eigendecomposition of $\varrho$ is given in \EQ{eq:rho_eigdecomp} and
\begin{equation}
\sigma=\sum_k \lambda_k \sigma_k.
\end{equation}
Clearly, such states given in \EQ{eq:rho12cc} form a convex set. Such states are called classical-quantum since in subsystem $1$ we have a mixture of states that are pairwise orthogonal to each other \cite{Bera2017Quantum}. We will denote the set of such states as $\mathcal C_1.$  For such couplings, the map given in \EQ{eq:map} needs only a von Neumann measurement.

Another possibility is the quantum-classical states, which  we will denote by $\mathcal C_2.$ Such states are of the form
\begin{equation}
\sum_k \lambda_k' \varrho_k \otimes \ketbra{k'},
\end{equation}
where the eigendecomposition of $\sigma$ is
\begin{equation}
\sigma=\sum_k \lambda'_k \ketbra{k'},
\end{equation}
and for the density matrices $\varrho_k$
\begin{equation}
\varrho=\sum_k \lambda'_k \varrho_k
\end{equation}
holds. A minimization over $\mathcal C_1$ or $\mathcal C_2$ will lead to a larger value than a minimization over separable couplings. An optimization over $\mathcal C_1$ or $\mathcal C_2$ can efficiently be carried out using semidefinite programming for any system size. We can also consider the set of states that are classical-classical, which are the members of both $\mathcal C_1$  and  $\mathcal C_2.$

Let us consider now the case of the GMPC self-distance, when $\varrho=\sigma.$ Then, a relevant coupling which is the elements of $\mathcal C_1$ and $\mathcal C_2$ is
\begin{equation}
\varrho_{cc}=\sum_k \lambda_k \ketbra{k} \otimes \ketbra{k},\label{eq:cc}
\end{equation}
where the eigendecomosition of $\varrho$ is given in \EQ{eq:rho_eigdecomp}. For the coupling in \EQ{eq:cc}, for $N=1$ the equals
\begin{equation}
\sum_k \lambda_k \va{H_1}_{\ket k} \le \va{H_1}_{\varrho}.
\end{equation}
So far, we have been obtaining results for the GMPC distance. Analogous statements hold for the DPT distance.

Another relevant case is the product state coupling
\begin{equation}
\varrho_{12}=\varrho \otimes \sigma.
\end{equation}
Then, we can define the distances given in \EQS{eq:GMPC_distance} and \eqref{eq:distance} for product states as
\begin{align}
&D_{\rm GMPC,prod}(\varrho,\sigma)^2=D_{\rm DPT,prod}(\varrho,\sigma)^2\nonumber\\
&\quad=\frac 1 2 \sum_{n=1}^N
\trace[(\Hamiltonian_n\otimes \openone-\openone\otimes \Hamiltonian_n)^2  \varrho \otimes \sigma ]\nonumber\\
&\quad=\frac1 2 \sum_{n=1}^N\left[\va{\Hamiltonian_n}_{\varrho}+\va{\Hamiltonian_n}_{\sigma}+(\ex{\Hamiltonian_n}_{\varrho}-\ex{\Hamiltonian_n}_{\sigma})^2\right],\nonumber\\
\end{align}
c.~f. \EQ{eq:eq:D2_pure_mixedB}. For the self-distance, the relation
\begin{equation}
D_{\rm GMPC,prod}(\varrho,\varrho)=D_{\rm DPT,prod}(\varrho,\varrho)=\sum_{n=1}^N\va{\Hamiltonian_n}_{\varrho}
\end{equation}
holds.

In Table~\ref{table:sets}, we summarized the self-distances obtained for the Wasserstein distance considering an optimization over various subsets of the bipartite quantum states and  $N=1.$ For the quantities in the Table, the inequality given in \EQ{eq:FQvar} holds. As expected, a minimization over a larger set will not lead to a larger value and often will lead to a smaller value.

\begin{table}[t!]
\begin{tabular}{|p{24mm}|p{23mm}|p{22mm}|}
\hline
Set of  \newline quantum states & GMPC \newline self-distance & DPT \newline self-distance\\
\hline
\hline
General & $\le {\mathcal F}_Q[\varrho,\Hamiltonian_1]/4$ & $I_\varrho(\Hamiltonian_1)$  \\
quantum states & &  \\
\hline
PPT states,& $\le {\mathcal F}_Q[\varrho,\Hamiltonian_1]/4$  & $\le {\mathcal F}_Q[\varrho,\Hamiltonian_1]/4$  \\
PPT$n$ states&  &   \\
\hline
Separable states & ${\mathcal F}_Q[\varrho,\Hamiltonian_1]/4$ & ${\mathcal F}_Q[\varrho,\Hamiltonian_1]/4$ \\
\hline
$\varrho_{cc}$ given  & $\sum_k \lambda_k \va{H_1}_{\ket k}$  & $\sum_k \lambda_k \va{H_1}_{\ket k}$ \\
in \EQ{eq:cc}& &  \\
\hline
Product state & $\va{\Hamiltonian_1}_{\varrho}$ & $\va{\Hamiltonian_1}_{\varrho}$ \\
\hline
\end{tabular}
\caption{Self-distance obtained for the Wasserstein distance considering an optimization over various subsets of the bipartite quantum states for $N=1.$} \label{table:sets}
\end{table}

We can consider an optimization over other convex sets of states. For instance, a convex set can be characterzied by constraints like
\begin{equation}
\va{A_k}\ge c_k
\end{equation}
or by the linear constraints
\begin{equation}
\ex{B_k}\ge d_k,
\end{equation}
where $A_k$ and $B_k$ are operators, $c_k$ and $d_k$ are constants. Other possibilities are the convex set of states with negativity not larger than a given bound \cite{Toth2018Quantum}, and the convex set of states not violating certain entanglement conditions. These conditions can be incorporated into the numerical optimization. We can also consider the convex set of states with a local hidden variable model \cite{Horodecki2009Quantum,Guhne2009Entanglement,Friis2019}.

\section{Alternative definition of the Wasserstein distance such that the self-distance equals various generalized quantum Fisher information quantities}
\label{sec:extension}

In this section, we will modify the definition of the Wasserstein distance such that the self-distance equals a quantity different from the ones we considered so far. In particular, we would like that it equals various generalized quantum Fisher information quantities. The Wigner-Yanase skew information and the quantum Fisher information are two members of this family.

The basic idea of \REFS{Petz2002Covariance,Gibilisco2009Quantum} is that for each standard matrix monotone function $f:\mathbb{R}^{+}\rightarrow\mathbb{R}^{+},$ a generalized variance and a corresponding quantum Fisher information are defined.  The notion standard means that  $f$ must satisfy
\begin{subequations}\begin{align}
f(1)&=1,\label{eq:cond1}\\
f(t)&=tf(t^{-1}).\label{eq:cond2}
\end{align}\end{subequations}
For a review on generalized variances, generalized quantum Fisher information quantities and covariances see \REF{Petz2011Introduction}. Moreover,  it is also useful to define the mean based on $f$ as
\begin{equation}
m_{f}(a,b)=af\left(\frac{b}{a}\right)
\end{equation}
 and use it instead of $f.$  The normalization condition given in  \EQ{eq:cond1} corresponds to  the condition
\begin{equation}
m_{f}(a,a)=a
\end{equation}
 for the means. The requirement given in \EQ{eq:cond2} corresponds to
 \begin{equation}
 m_{f}(a,b)=m_{f}(b,a).
 \end{equation}
A list of generalized quantum Fisher information quantities generated by various  well-known means $m_{f}(a,b)$ can be found in Refs.~\cite{Petz2002Covariance,Gibilisco2009Quantum}.

Using the normalization suggested by \REF{Toth2013Extremal}, we arrive at a family of generalized quantum Fisher information quantities $F^f_{Q}[\varrho,\Hamiltonian]$ such that (i) for pure states, we have
\begin{equation}
\mathcal F^f_{Q}[\ket{\Psi},\Hamiltonian]=4\va {\Hamiltonian}_{\Psi}.
\end{equation}
(ii) For mixed states, $F^f_{Q}[\varrho,\Hamiltonian]$ is convex in the state.

The generalized variance ${\rm var}^f_{\varrho}(\Hamiltonian)$ fulfills the following two requirements. (i) For pure states, the generalized variance equals the usual variance
\begin{equation}
{\rm var}^f_{\Psi}(\Hamiltonian)=\va{\Hamiltonian}_{\Psi}.
\end{equation}
(ii) For mixed states, ${\rm var}^f_{\varrho}(\Hamiltonian)$ is concave in the state.

A family of generalized quantum Fisher information and generalized variance fulfilling the above requirements are \cite{Toth2013Extremal}
\begin{subequations}
\begin{align}
{\mathcal F}_{Q}^{f}[\varrho,\Hamiltonian]&=2\sum_{k,l}\frac{m_{f}(1,0)}{m_{f}(\lambda_k,\lambda_l)}{(\lambda_k-\lambda_l)^{2}}\vert \Hamiltonian_{kl}\vert^{2},\label{eq:FQf}\\\nonumber
{\rm var}_{\varrho}^{f}(\Hamiltonian)&=\frac{1}{2}\sum_{k,l}\frac{m_{f}(\lambda_k,\lambda_l)}{m_{f}(1,0)}\vert \Hamiltonian_{kl}\vert^{2}\nonumber\\
&-\frac{1}{2m_{f}(1,0)}
\left\vert\sum\lambda_k\Hamiltonian_{kk}\right\vert^{2},\label{eq:varf}
\end{align}
\end{subequations}
where the matrix elements of $\Hamiltonian$ in the eigenbasis of $\varrho$ are denoted as
\begin{equation}
\Hamiltonian_{kl}=\langle k\vert \Hamiltonian\vert l\rangle.
\end{equation}
${\mathcal F}_{Q}^{f}[\varrho,\Hamiltonian]/4$ has been called metric-adjusted skew information \cite{Hansen2008Metric}.

Note that the definition of the variance in \EQ{eq:varf} requires that $m_{f}(1,0)\equiv f(0)$ is nonzero. In such cases $f$ is called regular \cite{Gibilisco2021AUnified}.

The usual quantum Fisher information and the usual variance corresponds to
\begin{equation}
f_{\max}(x)=\frac{1+x}{2},
\end{equation}
and the arithmetic mean
\begin{equation}
m_{f_{\max}}(a,b)=\frac{a+b}{2}.
\end{equation}
Note that $f_{\max}(x)$ is the largest among standard matrix monotone functions.
Due to this, ${\mathcal F}_{Q}^{f_{\max}}[\varrho,\Hamiltonian]$ is the largest among ${\mathcal F}_{Q}^{f}[\varrho,\Hamiltonian]$
and ${\rm var}_{\varrho}^{f_{\max}}(\Hamiltonian)$ is the smallest among ${\rm var}_{\varrho}^{f}(\Hamiltonian).$

The Wigner-Yanase skew information corresponds to
\begin{equation}
f_{\rm WY}(x)=\frac{(\sqrt x +1)^2}4,
\end{equation}
and the mean
\begin{equation}
m_{f_{\rm WY}}(a,b)=\frac{(\sqrt{a}+\sqrt{b})^2}{4}.
\end{equation}
Note that we get 4 times the usual Wigner-Yanase skew information due to the chosen normalization
\begin{equation}
\mathcal F^{f_{\rm WY}}_{Q}[\varrho,\Hamiltonian]=4I_{\varrho}(\Hamiltonian).
\end{equation}

We now show a method to express the various generalized quantum Fisher information quantities with each other.

\DEFOBS{obs:QFI_gen_X}Let us define for given $f$ the following matrix in the eigenbasis of $\varrho$
\begin{equation}
(X_f)_{kl}=\langle k \vert X\vert l \rangle=\sqrt{\frac{m_{f}(1,0)}{m_{f}(\lambda_k,\lambda_l)}}.
\end{equation}
Then, any generalized quantum Fisher information can be expressed as
\begin{equation}
{\mathcal F}_{Q}^{f_1}[\varrho,\Hamiltonian]={\mathcal F}^{f_2}_{Q}[\varrho,Q_{f_1,f_2} \circ \Hamiltonian],\label{eq:FQfX}
\end{equation}
where "$\circ$" denotes element-wise or Hadamard product defined as
\begin{equation}
A\circ B=\sum_{k,l} \langle k \vert A \vert l\rangle \langle k \vert B \vert l\rangle \ket{k}\bra{l},
\end{equation}
where $\ket{k}$ and $\ket{l}$ are the eigenvectors of the density matrix, and the coefficient for converting one type of quantum Fisher information into another ons is given as
\begin{equation}
(Q_{f_1,f_2})_{kl}=\begin{cases}\frac{(X_{f_1})_{kl}}{(X_{f_2})_{kl}}, &\text{if }\lambda_k\ne\lambda_l,\\
0, &\text{if }\lambda_k=\lambda_l.\end{cases}\label{eq:Qf1f2}
\end{equation}

 \PROOF The statement can be verified by direct comparison of the definition of the quantum Fisher information given in \EQS{eq:FQf} and \eqref{eq:FQfX}. $\qed$

It is interesting to compute the matrix needed for $f_1=f_{\max}$ and $f_2=f_{\rm WY}.$ We obtain, in the basis of the eigenvectors of $\varrho,$
\begin{equation}
(Q_{f_{\max},f_{\rm WY}})_{kl}=\begin{cases}\frac{(X_{f_{\max}})_{kl}}{(X_{f_{\rm WY}})_{kl}}=\frac{\sqrt{\lambda_k}+\sqrt{\lambda_l}}{\sqrt{\lambda_k+\lambda_l}}, &\text{if }\lambda_k\ne\lambda_l,\\
0, &\text{if }\lambda_k=\lambda_l.\end{cases}\label{eq:Q}
\end{equation}
Here note that in \EQ{eq:Q}, the denominator of the fraction with $\lambda_k$ and $\lambda_l$ is positive if $\lambda_k\ne\lambda_l.$ With the matrix in \EQ{eq:Q}, we can use the Wigner-Yanase skew information to obtain a the quantum Fisher information
\be
\mathcal F_Q[\varrho,\Hamiltonian]=4I_{\varrho}(Q_{f_{\max},f_{\rm WY}} \circ \Hamiltonian).
\end{equation}

Next, we compute the matrix needed for $f_1=f_{\rm WY}$ and $f_2=f_{\max}.$ We obtain, in the basis of the eigenvectors of $\varrho,$
\begin{equation}
(Q_{f_{\rm WY},f_{\max}})_{kl}=\begin{cases}\frac{(X_{f_{\rm WY}})_{kl}}{(X_{f_{\max}})_{kl}}=\frac{\sqrt{\lambda_k+\lambda_l}}{\sqrt{\lambda_k}+\sqrt{\lambda_l}}, &\text{if }\lambda_k\ne\lambda_l,\\
0, &\text{if }\lambda_k=\lambda_l.\end{cases}\label{eq:W}
\end{equation}
Then, we can also obtain the Wigner-Yanase skew information with the quantum Fisher information as
\begin{equation}
I_{\varrho}(\Hamiltonian)=\frac1 4 \mathcal F_Q[\varrho,Q_{f_{\rm WY},f_{\max}} \circ \Hamiltonian].
\end{equation}

Next, we show how the various generalized quantum Fisher information quantities can be expressed as a convex roof over the decompositions of the density matrix.

\DEFOBS{obs:QFI_f_conv_roof}The various generalized quantum Fisher information quantities can be expressed as a convex roof as
\begin{equation}
{\mathcal F}^f_Q[\varrho,\Hamiltonian]=4\min_{\{p_k,\ket{\Psi_k}\}} \sum_k p_k \vasq{(Y_f\circ \Hamiltonian)}_{\Psi_k},\label{eq:deffqroof_Xf}
\end{equation}
where the optimization is over pure state decompositions given in \EQ{eq:purestatedecomp}, and in the basis of the eigenvectors of $\varrho$ we define
\begin{align}
&(Y_f)_{kl}\nonumber\\
&\quad=\begin{cases}\frac{(X_f)_{kl}}{(X_{f_{\max}})_{kl}}=\sqrt{\frac{m_{f}(1,0)}{m_{f}(\lambda_k,\lambda_l)}(\lambda_k+\lambda_l)}, &\text{if }\lambda_k\ne\lambda_l,\\
0, &\text{if }\lambda_k=\lambda_l.\end{cases}\nonumber\\
\end{align}

 \PROOF It follows from \OBS{obs:QFI_gen_X} defining $(X_f)_{kl}$ and the definition of the quantum Fisher information with the convex roof of the variance given in \EQ{eq:deffqroof}. $\qed$

Next, we show how the various generalized quantum Fisher information quantities can be expressed as an optimization in the two-copy space.

\DEFOBS{obs:QFI_f_sep}The generalized quantum Fisher information can be obtained as an optimization over separable states
\begin{align}
{\mathcal F}_{Q}^{f}[\varrho,\Hamiltonian]=
\min_{ \varrho_{12} }\;&
2\trace[(Y_f\circ \Hamiltonian\otimes \openone-\openone\otimes Y_f\circ \Hamiltonian)^2  \varrho_{12} ],\nonumber\\
\textrm{s.~t. }&
\varrho_{12}\in \mathcal S,  \nonumber\\
&{\rm Tr}_2(\varrho_{12})=\varrho,  \nonumber\\
&{\rm Tr}_1(\varrho_{12})=\varrho.\label{eq:FQ_min_sep_states_Xf}
\end{align}

 \PROOF It follows from \OBS{obs:QFI_sep} defining the quantum Fisher information with an optimization over bipartite separable quantum states and \OBS{obs:QFI_gen_X} defining $(X_f)_{kl}.$ $\qed$

Let us calculate a concrete example. For instance, for obtaining the Wigner-Yanase skew information times four as an optimization over separable states, we should use
\begin{equation}
(Y_{f_{\rm WY}})_{kl}=(Q_{f_{\rm WY},f_{\max}})_{kl},
\end{equation}
where $Q_{f_{\rm WY},f_{\max}}$ is defined in \EQ{eq:W}.

\DEFOBS{obs:QFI_f_optimization_over_general_states}We can obtain the various quantum Fisher information quantities as an optimization over general quantum states, rather than over separable states, as
\begin{align}
{\mathcal F}_{Q}^{f}[\varrho,\Hamiltonian](\varrho)\nonumber\\
=\min_{ \varrho_{12} }\;&
2\trace\{[(Z_f\circ \Hamiltonian)^T\otimes \openone-\openone\otimes Z_f\circ \Hamiltonian]^2  \varrho_{12} \},\nonumber\\
\textrm{s.~t. }&
\varrho_{12}\in\mathcal D,\nonumber\\
& {\rm Tr}_2(\varrho_{12})=\varrho^T, \nonumber\\
& {\rm Tr}_1(\varrho_{12})=\varrho,\label{eq:QFI_f_optimization_over_general_states}
\end{align}
where in the basis of the eigenvectors of $\varrho$ we define
\begin{equation}
(Z_f)_{kl}=\begin{cases}\frac{(X_f)_{kl}}{(X_{f_{\rm WY}})_{kl}}, &\text{if }\lambda_k\ne\lambda_l,\\
0, &\text{if }\lambda_k=\lambda_l.\end{cases}\label{eq:Z}
\end{equation}

\PROOF We use the definition of $D_{\rm DPT}(\varrho,\sigma)^2$ given in \EQ{eq:distance} and the equation relating it to the Wigner-Yanase skew information given in \EQ{eq:DrhoI}, together with \EQ{eq:FQfX}. $\qed$

Let us again calculate a concrete example. For instance, for obtaining the quantum Fisher information as an optimization over general quantum states, we should use
\begin{equation}
(Z_{f_{\max}})_{kl}=Q_{f_{\max},f_{\rm WY}},
\end{equation}
where $Q_{f_{\max},f_{\rm WY}}$ is defined in \EQ{eq:Q}.

Based on these, we can define various Wasserstein distance measures, for which the self-distance equals various quantum Fisher information quantities.

\DEFDEFINITION{obs:D2_QFI_f_sep}A family of Wasserstein distance measures can be defined based on an optimization over separable states as
\begin{align}
D^f_{\rm GMPC,sep}(\varrho,\sigma)^2\nonumber\\
=\frac 1 2 \min_{ \varrho_{12} }\;&
\trace[(Y_f\circ \Hamiltonian\otimes \openone-\openone\otimes Y_f\circ \Hamiltonian)^2  \varrho_{12} ],\nonumber\\
\textrm{s.~t. }&
\varrho_{12}\in \mathcal S,  \nonumber\\
& {\rm Tr}_2(\varrho_{12})=\varrho, \nonumber\\
& {\rm Tr}_1(\varrho_{12})=\sigma.\label{eq:D2_FQ_min_sep_states_Xf}
\end{align}
Based on \OBS{obs:QFI_f_sep}, for the self-distance
\begin{equation}
D^f_{\rm GMPC,sep}(\varrho,\varrho)^2=\frac1 4 {\mathcal F}_{Q}^{f}[\varrho,\Hamiltonian]\label{eq:Df_GMP_sep_self_dist}
\end{equation}
holds.

\DEFDEFINITION{obs:D2_QFI_f_optimization_over_general_states}A family of Wasserstein distance measures can be defined based on an optimization over general quantum states as
\begin{align}
D^f_{\rm DPT}(\varrho,\sigma)^2\nonumber\\
=\frac 1 2
\min_{ \varrho_{12} }\;&
\trace\{[(Z_f\circ \Hamiltonian)^T\otimes \openone-\openone\otimes Z_f\circ \Hamiltonian]^2  \varrho_{12} \},\nonumber\\
\textrm{s.~t. }&
\varrho_{12}\in\mathcal D,\nonumber\\
& {\rm Tr}_2(\varrho_{12})=\varrho^T, \nonumber\\
& {\rm Tr}_1(\varrho_{12})=\sigma.\label{eq:D2_QFI_f_optimization_over_general_states}
\end{align}
Based on \OBS{obs:QFI_f_optimization_over_general_states}, for the self-distance
\begin{equation}
D^f_{\rm DPT}(\varrho,\varrho)^2=\frac1 4 {\mathcal F}_{Q}^{f}[\varrho,\Hamiltonian]\label{eq:Df_DPT_self_dist}
\end{equation}
holds. It would be interesting to examine the properties of the quantities defined in Definitions~\ref{obs:D2_QFI_f_sep} and \ref{obs:D2_QFI_f_optimization_over_general_states}.

\section{Conclusions}

We discussed how to define the quantum Wasserstein distance as an optimization over bipartite separable states rather than an optimization over general quantum states. With such a definition, the self-distance becomes related to the quantum Fisher information. We introduced also variance-like quantities in which we replaced the minimization used in the definition of the quantum Wasserstein distance by a maximization, and examined their properties. We discussed the relation of our findings to entanglement criteria. We examined also the quantity obtained after we considered optimizing the variance rather than the second moment in the usual expression of the quantum Wasserstein distance. Finally, we extended our results to the various generalized quantum Fisher information quantities. The details of the numerical calculations are discussed in \APP{app:Numerics}.

\begin{acknowledgments}
We thank I.~Apellaniz, M.~Eckstein, F. Fr\"owis, I.~L.~Egusquiza, C.~Klempt, J.~Ko\l ody\'nski, M.~W.~Mitchell, M.~Mosonyi, G.~Muga, J.~Siewert, Sz.~Szalay, K. \.Zyczkowski, G. Vitagliano, and D. Virosztek  for discussions. We acknowledge the support of the  EU (COST Action CA15220), the Spanish MCIU (Grant No. PCI2018-092896), the Spanish Ministry of Science, Innovation and Universities and the European Regional Development Fund FEDER through Grant No. PGC2018-101355-B-I00 (MCIU/AEI/FEDER, EU) and through Grant No. PID2021-126273NB-I00 funded by MCIN/AEI/10.13039/501100011033 and by "ERDF A way of making Europe", the Basque Government (Grant No. IT986-16, No. IT1470-22), and the National Research, Development and Innovation Office NKFIH (Grant No.  K124351, No. K124152, No. KH129601, 2019-2.1.7-ERA-NET-2022-00036).  We thank the "Frontline" Research Excellence Programme of the NKFIH (Grant No. KKP133827). This project has received funding from the European Union's Horizon 2020 Research and Innovation Programme under Grant Agreement no. 731473 and 101017733 (QuantERA MENTA, Quant\-ERA QuSiED).  We thank Project no. TKP2021-NVA-04, which has been implemented with the support provided by the Ministry of Innovation and Technology of Hungary from the National Research, Development and Innovation Fund, financed under the TKP2021-NVA funding scheme. We thank the Quantum Information National Laboratory of Hungary. G.~T. is thankful for a  Bessel Research Award from the Humboldt Foundation. J.~P. is partially supported by the Momentum program of the Hungarian Academy of Sciences under grant agreement no. LP2021-15/2021.
\end{acknowledgments}

\appendix

\section{Details of the numerical calculations}
\label{app:Numerics}

We used MATLAB \cite{MATLAB2020} for numerical calculations. We used the semidefinite solver MOSEK \cite{MOSEK} and the front-end YALMIP \cite{Lofberg2004Yalmip}. We also used the QUBIT4MATLAB package \cite{Toth2008QUBIT4MATLAB,QUBIT4MATLAB_actual_note62_href}. $D_{\rm DPT}(\varrho,\sigma)^2$ given in \DEFINITION{def:D} and $D_{\rm GMPC}(\varrho,\sigma)^2$ given in \DEFINITION{def:DGMCP} can be obtained using semidefinite programming.  $D_{\rm GMPC, sep}(\varrho,\sigma)^2$ in \DEFINITION{def:def1} and $D_{\rm DPT, sep}(\varrho,\sigma)^2$ in \DEFINITION{def:def2} need an optimization over separable states.  The optimization over separable states can be carried out numerically using semidefinite programming for two qubits, since in this case the set of PPT states equals the set of separable states.

We included the routines computing the various quantum Wasserstein distance measures as {\tt wdistsquare\_GMPC\_ppt.m},
{\tt wdistsquare\_GMPC.m},
{\tt wdistsquare\_DPT\_ppt.m}, and
{\tt wdistsquare\_DPT.m}.
We included the various types of the variance-like quantities as
{\tt wvar\_GMPC\_ppt.m},
{\tt wvar\_GMPC.m},
{\tt wvar\_DPT\_ppt.m}, and
{\tt wvar\_DPT.m}.
The usage of these routines is demonstrated in {\tt example\_wdistsquare.m}.

\bibliographystyle{quantum}
\bibliography{wasserstein_and_quantum_fisher68}

\begin{thebibliography}{100}

\bibitem{Monge1781Memoire}
{G. Monge}.
\newblock ``{M\'emoire sur la th\'eory des d\'eblais et des remblais}''.
\newblock {M\'emoires de l'Acad\'emie Royale de Sciences de Paris}~(1781).

\bibitem{Kantorovitch1958Translocation}
L.~Kantorovitch.
\newblock ``On the translocation of masses''.
\newblock Management Science {\bf 5}, 1--4~(1958).
\newblock
  url:~\href{http://www.jstor.org/stable/2626967}{http://www.jstor.org/stable/2626967}.

\bibitem{Boissard2015Distributions}
Emmanuel Boissard, Thibaut~Le Gouic, and Jean-Michel Loubes.
\newblock ``Distribution's template estimate with wasserstein metrics''.
\newblock \href{https://dx.doi.org/10.3150/13-bej585}{Bernoulli {\bf 21},
  740--759}~(2015).

\bibitem{Butkovsky2014Subgeometric}
Oleg Butkovsky.
\newblock ``{Subgeometric rates of convergence of Markov processes in the
  Wasserstein metric}''.
\newblock \href{https://dx.doi.org/10.1214/13-AAP922}{Ann. Appl. Probab. {\bf
  24}, 526--552}~(2014).

\bibitem{Hairer2011Asymptotic}
{M. Hairer, J.-C. Mattingly and M. Scheutzow}.
\newblock ``{Asymptotic coupling and a general form of Harris' theorem with
  applications to stochastic delay equations}''.
\newblock \href{https://dx.doi.org/10.1007/s00440-009-0250-6}{Probab. Theory
  Relat. Fields {\bf 149}, 223--259}~(2011).

\bibitem{Hairer2008Spectral}
M.~Hairer and J.C. Mattingly.
\newblock ``{Spectral Gaps in Wasserstein Distances and the 2D Stochastic
  Navier-Stokes Equations}''.
\newblock \href{https://dx.doi.org/10.1214/08-AOP392}{Ann. Probab. {\bf 36},
  2050--2091}~(2008).

\bibitem{Figalli2010Mass}
{A. Figalli, F. Maggi and A. Pratelli}.
\newblock ``A mass transportation approach to quantitative isoperimetric
  inequalities''.
\newblock \href{https://dx.doi.org/10.1007/s00222-010-0261-z}{Invent. Math.
  {\bf 182}, 167--211.}~(2010).

\bibitem{Figalli2011Shape}
A.~Figalli and F.~Maggi.
\newblock ``On the shape of liquid drops and crystals in the small mass
  regime''.
\newblock \href{https://dx.doi.org/10.1007/s00205-010-0383-x}{Arch. Ration.
  Mech. Anal. {\bf 201}, 143--207}~(2011).

\bibitem{LottVillani2009Ricci}
J.~Lott and C.~Villani.
\newblock ``{Ricci curvature for metric-measure spaces via optimal
  transport}''.
\newblock \href{https://dx.doi.org/10.48550/arXiv.math/0412127}{{Ann. of Math.}
  {\bf 169 (3)}, 903--991}~(2009).

\bibitem{RenesseSturm2005Transport}
{Max-K. von Renesse and Karl-Theodor Sturm}.
\newblock ``{Transport inequalities, gradient estimates, entropy, and Ricci
  curvature}''.
\newblock \href{https://dx.doi.org/10.1002/cpa.20060}{Comm. Pure Appl. Math.
  {\bf 58}, 923--940}~(2005).

\bibitem{Sturm2006Geometry}
Karl-Theodor Sturm.
\newblock ``{On the geometry of metric measure spaces I}''.
\newblock \href{https://dx.doi.org/10.1007/s11511-006-0002-8}{Acta Math. {\bf
  196}, 65--131}~(2006).

\bibitem{Sturm2006Geometry2}
Karl-Theodor Sturm.
\newblock ``{On the geometry of metric measure spaces II}''.
\newblock \href{https://dx.doi.org/10.1007/s11511-006-0003-7}{Acta Math. {\bf
  196}, 133--177}~(2006).

\bibitem{Kloeckner2010Geometric}
{Beno\^{\i}t Kloeckner}.
\newblock ``{A geometric study of Wasserstein spaces: Euclidean spaces}''.
\newblock \href{https://dx.doi.org/10.2422/2036-2145.2010.2.03}{{Annali della
  Scuola Normale Superiore di Pisa - Classe di Scienze, Scuola Normale
  Superiore 2010} {\bf IX (2)}, 297--323}~(2010).

\bibitem{Geherr2019}
Gy\"orgy~P\'al Geh\'er, Tam\'as Titkos, and D\'aniel Virosztek.
\newblock ``On isometric embeddings of wasserstein spaces -- the discrete
  case''.
\newblock
  \href{https://dx.doi.org/https://doi.org/10.1016/j.jmaa.2019.123435}{J. Math.
  Anal. Appl. {\bf 480}, 123435}~(2019).

\bibitem{Viro2020}
{Gy\"orgy P\'al Geh\'er, T. Titkos, D\'aniel Virosztek}.
\newblock ``{Isometric study of Wasserstein spaces -- the real line}''.
\newblock \href{https://dx.doi.org/10.1090/tran/8113}{{Trans. Amer. Math. Soc.}
  {\bf 373}, 5855--5883}~(2020).

\bibitem{Viro2021}
Gy\"orgy~P\'al Geh\'er, Tam\'as Titkos, and D\'aniel Virosztek.
\newblock ``{The isometry group of Wasserstein spaces: the Hilbertian case}''.
\newblock \href{https://dx.doi.org/10.1112/jlms.12676}{J. Lond. Math. Soc. {\bf
  106}, 3865--3894}~(2022).

\bibitem{Viro2022}
Gy\"orgy~P\'al Geh\'er, Tam\'as Titkos, and D\'aniel Virosztek.
\newblock ``Isometric rigidity of wasserstein tori and spheres''.
\newblock \href{https://dx.doi.org/10.1112/mtk.12174}{Mathematika {\bf 69},
  20--32}~(2023).

\bibitem{Kiss2022}
Gergely Kiss and Tam{\'{a}}s Titkos.
\newblock ``Isometric rigidity of wasserstein spaces: The graph metric case''.
\newblock \href{https://dx.doi.org/10.1090/proc/15977}{Proc. Am. Math. Soc.
  {\bf 150}, 4083--4097}~(2022).

\bibitem{Geher2023}
Gy\"orgy~P\'al Geh\'er, Tam\'as Titkos, and D\'aniel Virosztek.
\newblock ``On the exotic isometry flow of the quadratic wasserstein space over
  the real line''.
\newblock
  \href{https://dx.doi.org/https://doi.org/10.1016/j.laa.2023.02.016}{Linear
  Algebra Appl.}~(2023).

\bibitem{Kolouri2016Radon}
{S. Kolouri, S.~R. Park and G.~K. Rohde}.
\newblock ``{The Radon cumulative distribution transform and its application to
  image classification}''.
\newblock \href{https://dx.doi.org/10.1109/TIP.2015.2509419}{IEEE Trans. Image
  Process. {\bf 25}, 920--934}~(2016).

\bibitem{Wang2013Linear}
{W. Wang, D. Slep\u{c}ev, S. Basu, J.~A. Ozolek and G.~K. Rohde}.
\newblock ``{A linear optimal transportation framework for quantifying and
  visualizing variations in sets of images}''.
\newblock \href{https://dx.doi.org/10.1007/s11263-012-0566-z}{{Int. J. Comput.
  Vis.} {\bf 101}, 254--269}~(2013).

\bibitem{Kolouri2017Optimal}
{S. Kolouri, S. Park, M. Thorpe, D. Slep\u{c}ev, G.~K. Rohde}.
\newblock ``{Optimal Mass Transport: Signal processing and machine-learning
  applications}''.
\newblock \href{https://dx.doi.org/10.1109/MSP.2017.2695801}{{IEEE Signal
  Processing Magazine} {\bf 34}, 43--59}~(2017).

\bibitem{Gramfort2015Fast}
{A. Gramfort, G. Peyr\'e and M. Cuturi}.
\newblock ``{Fast Optimal Transport Averaging of Neuroimaging Data}''.
\newblock \href{https://dx.doi.org/10.1007/978-3-319-19992-4_20}{{Information
  Processing in Medical Imaging. IPMI 2015. Lecture Notes in Computer Science}
  {\bf 9123}, 261--272}~(2015).

\bibitem{Su2015Shape}
{Z. Su, W. Zeng, Y. Wang, Z.~L. Lu and X. Gu}.
\newblock ``{Shape classification using Wasserstein distance for brain
  morphometry analysis}''.
\newblock \href{https://dx.doi.org/10.1007/978-3-319-19992-4_32}{{Information
  Processing in Medical Imaging. IPMI 2015. Lecture Notes in Computer Science}
  {\bf 24}, 411--423}~(2015).

\bibitem{Arjovsky2017Wasserstein}
Martin Arjovsky, Soumith Chintala, and L{\'e}on Bottou.
\newblock ``{W}asserstein generative adversarial networks''.
\newblock In Doina Precup and Yee~Whye Teh, editors, Proceedings of the 34th
  International Conference on Machine Learning.
\newblock Volume~70 of Proceedings of Machine Learning Research, pages
  214--223.
\newblock PMLR~(2017).
\newblock  \href{http://arxiv.org/abs/1701.07875}{arXiv:1701.07875}.

\bibitem{Moselhy2012Bayesian}
{T.~A. El Moselhy and Y.~M. Marzouk}.
\newblock ``{Bayesian inference with optimal maps}''.
\newblock \href{https://dx.doi.org/10.1016/j.jcp.2012.07.022}{{J. Comput.
  Phys.} {\bf 231}, 7815--7850}~(2012).

\bibitem{Peyre2019Computational}
{Gabriel Peyr\'e and Marco Cuturi}.
\newblock ``{Computational Optimal Transport: With Applications to Data
  Science}''.
\newblock \href{https://dx.doi.org/10.1561/2200000073}{{Found. Trends Machine
  Learn.} {\bf 11}, 355--602}~(2019).

\bibitem{Frogner2015Learning}
Charlie Frogner, Chiyuan Zhang, Hossein Mobahi, Mauricio Araya, and Tomaso~A
  Poggio.
\newblock ``Learning with a wasserstein loss''.
\newblock In C.~Cortes, N.~Lawrence, D.~Lee, M.~Sugiyama, and R.~Garnett,
  editors, Advances in Neural Information Processing Systems.
\newblock Volume~28.
\newblock Curran Associates, Inc.~(2015).
\newblock  \href{http://arxiv.org/abs/1506.05439}{arXiv:1506.05439}.

\bibitem{Ramdas2017Wasserstein}
{A. Ramdas, N. G. Trillos and M. Cuturi}.
\newblock ``{On Wasserstein Two-Sample Testing and Related Families of
  Nonparametric Tests}''.
\newblock \href{https://dx.doi.org/10.3390/e19020047}{{Entropy} {\bf 19},
  47.}~(2017).

\bibitem{Srivastava2018Scalable}
{S. Srivastava, C. Li and D. B. Dunson}.
\newblock ``{Scalable Bayes via Barycenter in Wasserstein Space}''.
\newblock {J. Mach. Learn. Res.} {\bf 19}, 1--35~(2018).
\newblock  \href{http://arxiv.org/abs/1508.05880}{arXiv:1508.05880}.

\bibitem{Zyczkowski1998TheMonge}
Karol \.Zyczkowski and Wojeciech Slomczynski.
\newblock ``The {Monge} distance between quantum states''.
\newblock \href{https://dx.doi.org/10.1088/0305-4470/31/45/009}{J. Phys. A:
  Math. Gen. {\bf 31}, 9095--9104}~(1998).

\bibitem{Zyczkowski2001TheMonge}
Karol \.Zyczkowski and Wojciech Slomczynski.
\newblock ``The {Monge} metric on the sphere and geometry of quantum states''.
\newblock \href{https://dx.doi.org/10.1088/0305-4470/34/34/311}{J. Phys. A:
  Math. Gen. {\bf 34}, 6689--6722}~(2001).

\bibitem{Bengtsson2006Geometry}
Ingemar Bengtsson and Karol \.Zyczkowski.
\newblock ``Geometry of quantum states: An introduction to quantum
  entanglement''.
\newblock \href{https://dx.doi.org/10.1017/CBO9780511535048}{Cambridge
  University Press}. ~(2006).

\bibitem{Biane2011Free}
{P. Biane and D. Voiculescu}.
\newblock ``{A free probability analogue of the Wasserstein metric on the
  trace-state space}''.
\newblock \href{https://dx.doi.org/10.1007/s00039-001-8226-4}{GAFA, Geom.
  Funct. Anal. {\bf 11}, 1125--1138}~(2001).

\bibitem{CarlenMaas2014Analog}
Eric~A. Carlen and Jan Maas.
\newblock ``{An Analog of the 2-Wasserstein Metric in Non-Commutative
  Probability Under Which the Fermionic Fokker-Planck Equation is Gradient Flow
  for the Entropy}''.
\newblock \href{https://dx.doi.org/10.1007/s00220-014-2124-8}{Commun. Math.
  Phys. {\bf 331}, 887--926}~(2014).

\bibitem{CarlenMaas2017Gradient}
Eric~A. Carlen and Jan Maas.
\newblock ``{Gradient flow and entropy inequalities for quantum Markov
  semigroups with detailed balance}''.
\newblock \href{https://dx.doi.org/10.1016/j.jfa.2017.05.003}{J. Funct. Anal.
  {\bf 273}, 1810--1869}~(2017).

\bibitem{CarlenMaas2020Non-commutative}
Eric~A. Carlen and Jan Maas.
\newblock ``Non-commutative calculus, optimal transport and functional
  inequalities in dissipative quantum systems''.
\newblock \href{https://dx.doi.org/10.1007/s10955-019-02434-w}{J. Stat. Phys.
  {\bf 178}, 319--378}~(2020).

\bibitem{DattaRouze2019Concentration}
Nilanjana Datta and Cambyse Rouz\'e.
\newblock ``Concentration of quantum states from quantum functional and
  transportation cost inequalities''.
\newblock \href{https://dx.doi.org/10.1063/1.5023210}{J. Math. Phys. {\bf 60},
  012202}~(2019).

\bibitem{DattaRouze2020Relating}
Nilanjana Datta and Cambyse Rouz\'e.
\newblock ``{Relating relative entropy, optimal transport and Fisher
  information: A quantum HWI inequality}''.
\newblock \href{https://dx.doi.org/10.1007/s00023-020-00891-8}{Ann. Henri
  Poincar\'e {\bf 21}, 2115--2150}~(2020).

\bibitem{Golse2016On}
Fran{\c{c}}ois Golse, Cl{\'e}ment Mouhot, and Thierry Paul.
\newblock ``On the mean field and classical limits of quantum mechanics''.
\newblock \href{https://dx.doi.org/10.1007/s00220-015-2485-7}{Commun. Math.
  Phys. {\bf 343}, 165--205}~(2016).

\bibitem{Golse2017The}
Fran{\c{c}}ois Golse and Thierry Paul.
\newblock ``The {Schr{\"o}dinger} equation in the mean-field and semiclassical
  regime''.
\newblock \href{https://dx.doi.org/10.1007/s00205-016-1031-x}{Arch. Ration.
  Mech. Anal. {\bf 223}, 57--94}~(2017).

\bibitem{Golse2018Wave}
Fran{\c{c}}ois Golse and Thierry Paul.
\newblock ``Wave packets and the quadratic {Monge}-{Kantorovich} distance in
  quantum mechanics''.
\newblock
  \href{https://dx.doi.org/https://doi.org/10.1016/j.crma.2017.12.007}{Comptes
  Rendus Math. {\bf 356}, 177--197}~(2018).

\bibitem{Golse2018TheQuantum}
Fran{\c{c}}ois Golse.
\newblock ``The quantum {$N$}-body problem in the mean-field and semiclassical
  regime''.
\newblock \href{https://dx.doi.org/10.1098/rsta.2017.0229}{Phil. Trans. R. Soc.
  A {\bf 376}, 20170229}~(2018).

\bibitem{Caglioti2020Quantum}
E.~Caglioti, F.~Golse, and T.~Paul.
\newblock ``Quantum optimal transport is cheaper''.
\newblock \href{https://dx.doi.org/10.1007/s10955-020-02571-7}{J. Stat. Phys.
  {\bf 181}, 149--162}~(2020).

\bibitem{Caglioti2021Towards}
Emanuele Caglioti, Fran{\c{c}}ois Golse, and Thierry Paul.
\newblock ``Towards optimal transport for quantum densities''.
\newblock
  \href{https://dx.doi.org/10.48550/arXiv.2101.03256}{{arXiv:2101.03256}}~(2021).

\bibitem{DePalma2021Quantum}
Giacomo De~Palma and Dario Trevisan.
\newblock ``Quantum optimal transport with quantum channels''.
\newblock \href{https://dx.doi.org/10.1007/s00023-021-01042-3}{Ann. Henri
  Poincar{\'e} {\bf 22}, 3199--3234}~(2021).

\bibitem{DePalma2021TheQuantum}
Giacomo De~Palma, Milad Marvian, Dario Trevisan, and Seth Lloyd.
\newblock ``The quantum {Wasserstein} distance of order 1''.
\newblock \href{https://dx.doi.org/10.1109/TIT.2021.3076442}{IEEE Trans. Inf.
  Theory {\bf 67}, 6627--6643}~(2021).

\bibitem{Friedland2022Quantum}
{Shmuel Friedland, Micha{\l} Eckstein, Sam Cole, and Karol \.Zyczkowski}.
\newblock ``{Quantum Monge--Kantorovich problem and transport distance between
  density matrices}''.
\newblock \href{https://dx.doi.org/10.1103/PhysRevLett.129.110402}{{Phys. Rev.
  Lett.} {\bf 129}, 110402}~(2022).

\bibitem{Friedland2021Quantum}
{Sam Cole, Micha{\l} Eckstein, Shmuel Friedland, and Karol \.Zyczkowski}.
\newblock ``Quantum optimal transport''.
\newblock
  \href{https://dx.doi.org/10.48550/arXiv.2105.06922}{{arXiv:2105.06922}}~(2021).

\bibitem{Bistron2022Monotonicity}
R.~Bistro\'n, M.~Eckstein, and K.~\.Zyczkowski.
\newblock ``{Monotonicity of a quantum 2-Wasserstein distance}''.
\newblock \href{https://dx.doi.org/10.1088/1751-8121/acb9c8}{J. Phys. A: Math.
  Theor. {\bf 56}, 095301}~(2023).

\bibitem{Geher2023Quantum}
Gy\"orgy~P\'al Geh\'er, J\'ozsef Pitrik, Tam\'as Titkos, and D\'aniel
  Virosztek.
\newblock ``{Quantum Wasserstein isometries on the qubit state space}''.
\newblock
  \href{https://dx.doi.org/https://doi.org/10.1016/j.jmaa.2022.126955}{J. Math.
  Anal. Appl. {\bf 522}, 126955}~(2023).

\bibitem{Li2022Wasserstein}
Lu~Li, Kaifeng Bu, Dax~Enshan Koh, Arthur Jaffe, and Seth Lloyd.
\newblock ``Wasserstein complexity of quantum circuits''.
\newblock \href{https://dx.doi.org/10.48550/arXiv.2208.06306}{{arXiv:
  2208.06306}}~(2022).

\bibitem{Kiani2022Learning}
Bobak~Toussi Kiani, Giacomo~De Palma, Milad Marvian, Zi-Wen Liu, and Seth
  Lloyd.
\newblock ``Learning quantum data with the quantum earth mover's distance''.
\newblock \href{https://dx.doi.org/10.1088/2058-9565/ac79c9}{Quantum Sci.
  Technol. {\bf 7}, 045002}~(2022).

\bibitem{Wigner1963INFORMATION}
E.~P. Wigner and Mutsuo~M. Yanase.
\newblock ``Information contents of distributions''.
\newblock \href{https://dx.doi.org/10.1073/pnas.49.6.910}{Proc. Natl. Acad.
  Sci. U.S.A. {\bf 49}, 910--918}~(1963).

\bibitem{Horodecki2009Quantum}
Ryszard Horodecki, Pawe\l{} Horodecki, Micha\l{} Horodecki, and Karol
  Horodecki.
\newblock ``Quantum entanglement''.
\newblock \href{https://dx.doi.org/10.1103/RevModPhys.81.865}{Rev. Mod. Phys.
  {\bf 81}, 865--942}~(2009).

\bibitem{Guhne2009Entanglement}
Otfried G{\"u}hne and G{\'e}za T{\'o}th.
\newblock ``Entanglement detection''.
\newblock
  \href{https://dx.doi.org/https://doi.org/10.1016/j.physrep.2009.02.004}{Phys.
  Rep. {\bf 474}, 1--75}~(2009).

\bibitem{Friis2019}
Nicolai Friis, Giuseppe Vitagliano, Mehul Malik, and Marcus Huber.
\newblock ``Entanglement certification from theory to experiment''.
\newblock \href{https://dx.doi.org/10.1038/s42254-018-0003-5}{Nat. Rev. Phys.
  {\bf 1}, 72--87}~(2019).

\bibitem{Giovannetti2004Quantum-Enhanced}
Vittorio Giovannetti, Seth Lloyd, and Lorenzo Maccone.
\newblock ``Quantum-enhanced measurements: Beating the standard quantum
  limit''.
\newblock \href{https://dx.doi.org/10.1126/science.1104149}{Science {\bf 306},
  1330--1336}~(2004).

\bibitem{Paris2009QUANTUM}
Matteo G.~A. Paris.
\newblock ``Quantum estimation for quantum technology''.
\newblock \href{https://dx.doi.org/10.1142/S0219749909004839}{Int. J. Quant.
  Inf. {\bf 07}, 125--137}~(2009).

\bibitem{Demkowicz-Dobrzanski2014Quantum}
Rafal Demkowicz-Dobrzanski, Marcin Jarzyna, and Jan Kolodynski.
\newblock ``Chapter four - {Quantum} limits in optical interferometry''.
\newblock \href{https://dx.doi.org/10.1016/bs.po.2015.02.003}{Prog. Optics {\bf
  60}, 345 -- 435}~(2015).
\newblock  \href{http://arxiv.org/abs/1405.7703}{arXiv:1405.7703}.

\bibitem{Pezze2014Quantum}
Luca Pezze and Augusto Smerzi.
\newblock ``Quantum theory of phase estimation''.
\newblock In G.M. Tino and M.A. Kasevich, editors, Atom Interferometry (Proc.
  Int. School of Physics 'Enrico Fermi', Course 188, Varenna).
\newblock Pages 691--741.
\newblock IOS Press, Amsterdam~(2014).
\newblock  \href{http://arxiv.org/abs/1411.5164}{arXiv:1411.5164}.

\bibitem{Toth2013Extremal}
G\'eza T\'oth and D\'enes Petz.
\newblock ``{Extremal properties of the variance and the quantum Fisher
  information}''.
\newblock \href{https://dx.doi.org/10.1103/PhysRevA.87.032324}{Phys. Rev. A
  {\bf 87}, 032324}~(2013).

\bibitem{Yu2013Quantum}
Sixia Yu.
\newblock ``{Quantum Fisher Information as the Convex Roof of Variance}''.
\newblock
  \href{https://dx.doi.org/10.48550/arXiv.1302.5311}{{arXiv:1302.5311}}~(2013).

\bibitem{Toth2022Uncertainty}
G\'eza T\'oth and Florian Fr\"owis.
\newblock ``{Uncertainty relations with the variance and the quantum Fisher
  information based on convex decompositions of density matrices}''.
\newblock \href{https://dx.doi.org/10.1103/PhysRevResearch.4.013075}{Phys. Rev.
  Research {\bf 4}, 013075}~(2022).

\bibitem{Chiew2022Improving}
Shao-Hen Chiew and Manuel Gessner.
\newblock ``Improving sum uncertainty relations with the quantum {Fisher}
  information''.
\newblock \href{https://dx.doi.org/10.1103/PhysRevResearch.4.013076}{Phys. Rev.
  Research {\bf 4}, 013076}~(2022).

\bibitem{Helstrom1976Quantum}
C.~W. Helstrom.
\newblock ``Quantum detection and estimation theory''.
\newblock Academic Press, New York. ~(1976).
\newblock
  url:~\href{https://www.elsevier.com/books/quantum-detection-and-estimation-theory/helstrom/978-0-12-340050-5}{www.elsevier.com/books/quantum-detection-and-estimation-theory/helstrom/978-0-12-340050-5}.

\bibitem{Holevo1982Probabilistic}
A.~S. Holevo.
\newblock ``Probabilistic and statistical aspects of quantum theory''.
\newblock North-Holland, Amsterdam. ~(1982).

\bibitem{Braunstein1994Statistical}
Samuel~L. Braunstein and Carlton~M. Caves.
\newblock ``Statistical distance and the geometry of quantum states''.
\newblock \href{https://dx.doi.org/10.1103/PhysRevLett.72.3439}{Phys. Rev.
  Lett. {\bf 72}, 3439--3443}~(1994).

\bibitem{Braunstein1996Generalized}
Samuel~L Braunstein, Carlton~M Caves, and Gerard~J Milburn.
\newblock ``Generalized uncertainty relations: Theory, examples, and {Lorentz}
  invariance''.
\newblock \href{https://dx.doi.org/10.1006/aphy.1996.0040}{Ann. Phys. {\bf
  247}, 135--173}~(1996).

\bibitem{Petz2008Quantum}
D{\'e}nes Petz.
\newblock ``Quantum information theory and quantum statistics''.
\newblock \href{https://dx.doi.org/10.1007/978-3-540-74636-2}{Springer, Berlin,
  Heilderberg}. ~(2008).

\bibitem{Toth2014Quantum}
G\'eza T\'oth and Iagoba Apellaniz.
\newblock ``Quantum metrology from a quantum information science perspective''.
\newblock \href{https://dx.doi.org/10.1088/1751-8113/47/42/424006}{J. Phys. A:
  Math. Theor. {\bf 47}, 424006}~(2014).

\bibitem{Pezze2018Quantum}
Luca Pezz\`e, Augusto Smerzi, Markus~K. Oberthaler, Roman Schmied, and Philipp
  Treutlein.
\newblock ``Quantum metrology with nonclassical states of atomic ensembles''.
\newblock \href{https://dx.doi.org/10.1103/RevModPhys.90.035005}{Rev. Mod.
  Phys. {\bf 90}, 035005}~(2018).

\bibitem{Barbieri2022Optical}
Marco Barbieri.
\newblock ``Optical quantum metrology''.
\newblock \href{https://dx.doi.org/10.1103/PRXQuantum.3.010202}{PRX Quantum
  {\bf 3}, 010202}~(2022).

\bibitem{Leka2013Some_B}
Zolt{\'a}n L{\'e}ka and D{\'e}nes Petz.
\newblock ``Some decompositions of matrix variances''.
\newblock Probab. Math. Statist. {\bf 33}, 191--199~(2013).
\newblock  \href{http://arxiv.org/abs/1408.2707}{arXiv:1408.2707}.

\bibitem{Petz2014}
D{\'e}nes Petz and D{\'a}niel Virosztek.
\newblock ``A characterization theorem for matrix variances''.
\newblock \href{https://dx.doi.org/10.14232/actasm-013-789-z}{Acta Sci. Math.
  (Szeged) {\bf 80}, 681--687}~(2014).

\bibitem{Fujiwara2008A}
Akio Fujiwara and Hiroshi Imai.
\newblock ``A fibre bundle over manifolds of quantum channels and its
  application to quantum statistics''.
\newblock \href{https://dx.doi.org/10.1088/1751-8113/41/25/255304}{J. Phys. A:
  Math. Theor. {\bf 41}, 255304}~(2008).

\bibitem{Escher2011General}
B.~M. Escher, R.~L. de~Matos~Filho, and L.~Davidovich.
\newblock ``General framework for estimating the ultimate precision limit in
  noisy quantum-enhanced metrology''.
\newblock \href{https://dx.doi.org/10.1038/nphys1958}{Nat. Phys. {\bf 7},
  406--411}~(2011).

\bibitem{Demkowicz-Dobrzanski2012The}
Rafa{\l} Demkowicz-Dobrza{\'n}ski, Jan Ko{\l}ody{\'n}ski, and
  M{\u{a}}d{\u{a}}lin Gu{\c{t}}{\u{a}}.
\newblock ``The elusive {Heisenberg} limit in quantum-enhanced metrology''.
\newblock \href{https://dx.doi.org/10.1038/ncomms2067}{Nat. Commun. {\bf 3},
  1063}~(2012).

\bibitem{Marvian2022Operational}
Iman Marvian.
\newblock ``Operational interpretation of quantum fisher information in quantum
  thermodynamics''.
\newblock \href{https://dx.doi.org/10.1103/PhysRevLett.129.190502}{Phys. Rev.
  Lett. {\bf 129}, 190502}~(2022).

\bibitem{Werner1989Quantum}
Reinhard~F. Werner.
\newblock ``{Quantum states with Einstein-Podolsky-Rosen correlations admitting
  a hidden-variable model}''.
\newblock \href{https://dx.doi.org/10.1103/PhysRevA.40.4277}{Phys. Rev. A {\bf
  40}, 4277--4281}~(1989).

\bibitem{Eckert2002Quantum}
K.~Eckert, J.~Schliemann, D.~Bruss, and M.~Lewenstein.
\newblock ``Quantum correlations in systems of indistinguishable particles''.
\newblock \href{https://dx.doi.org/10.1006/aphy.2002.6268}{Ann. Phys. {\bf
  299}, 88--127}~(2002).

\bibitem{Ichikawa2008Exchange}
Tsubasa Ichikawa, Toshihiko Sasaki, Izumi Tsutsui, and Nobuhiro Yonezawa.
\newblock ``Exchange symmetry and multipartite entanglement''.
\newblock \href{https://dx.doi.org/10.1103/PhysRevA.78.052105}{Phys. Rev. A
  {\bf 78}, 052105}~(2008).

\bibitem{Horodecki1997Separability}
Pawel Horodecki.
\newblock ``Separability criterion and inseparable mixed states with positive
  partial transposition''.
\newblock \href{https://dx.doi.org/10.1016/S0375-9601(97)00416-7}{Phys. Lett. A
  {\bf 232}, 333--339}~(1997).

\bibitem{Peres1996Separability}
Asher Peres.
\newblock ``Separability criterion for density matrices''.
\newblock \href{https://dx.doi.org/10.1103/PhysRevLett.77.1413}{Phys. Rev.
  Lett. {\bf 77}, 1413--1415}~(1996).

\bibitem{Horodecki1999Bound}
Pawe\l{} Horodecki, Micha\l{} Horodecki, and Ryszard Horodecki.
\newblock ``Bound entanglement can be activated''.
\newblock \href{https://dx.doi.org/10.1103/PhysRevLett.82.1056}{Phys. Rev.
  Lett. {\bf 82}, 1056--1059}~(1999).

\bibitem{Toth2018Quantum}
G\'eza T\'oth and Tam\'as V\'ertesi.
\newblock ``Quantum states with a positive partial transpose are useful for
  metrology''.
\newblock \href{https://dx.doi.org/10.1103/PhysRevLett.120.020506}{Phys. Rev.
  Lett. {\bf 120}, 020506}~(2018).

\bibitem{Hill1997Entanglement}
Scott Hill and William~K. Wootters.
\newblock ``Entanglement of a pair of quantum bits''.
\newblock \href{https://dx.doi.org/10.1103/PhysRevLett.78.5022}{Phys. Rev.
  Lett. {\bf 78}, 5022--5025}~(1997).

\bibitem{Wootters1998Entanglement}
William~K. Wootters.
\newblock ``Entanglement of formation of an arbitrary state of two qubits''.
\newblock \href{https://dx.doi.org/10.1103/PhysRevLett.80.2245}{Phys. Rev.
  Lett. {\bf 80}, 2245--2248}~(1998).

\bibitem{DiVincenzo1999Proceedings}
David~P. DiVincenzo, Christopher~A. Fuchs, Hideo Mabuchi, John~A. Smolin,
  Ashish Thapliyal, and Armin Uhlmann.
\newblock ``Entanglement of assistance''.
\newblock
  \href{https://dx.doi.org/10.48550/arXiv.quant-ph/9803033}{{quant-ph/9803033}}~(1998).

\bibitem{Smolin2005Entanglement}
John~A. Smolin, Frank Verstraete, and Andreas Winter.
\newblock ``Entanglement of assistance and multipartite state distillation''.
\newblock \href{https://dx.doi.org/10.1103/PhysRevA.72.052317}{Phys. Rev. A
  {\bf 72}, 052317}~(2005).

\bibitem{Hofmann2003Violation}
Holger~F. Hofmann and Shigeki Takeuchi.
\newblock ``Violation of local uncertainty relations as a signature of
  entanglement''.
\newblock \href{https://dx.doi.org/10.1103/PhysRevA.68.032103}{Phys. Rev. A
  {\bf 68}, 032103}~(2003).

\bibitem{Guhne2004Characterizing}
Otfried G\"uhne.
\newblock ``Characterizing entanglement via uncertainty relations''.
\newblock \href{https://dx.doi.org/10.1103/PhysRevLett.92.117903}{Phys. Rev.
  Lett. {\bf 92}, 117903}~(2004).

\bibitem{Guhne2006Entanglement}
Otfried G\"uhne, M\'aty\'as Mechler, G\'eza T\'oth, and Peter Adam.
\newblock ``Entanglement criteria based on local uncertainty relations are
  strictly stronger than the computable cross norm criterion''.
\newblock \href{https://dx.doi.org/10.1103/PhysRevA.74.010301}{Phys. Rev. A
  {\bf 74}, 010301}~(2006).

\bibitem{Vitagliano2011Spin}
Giuseppe Vitagliano, Philipp Hyllus, I{\~n}igo~L. Egusquiza, and G\'eza T\'oth.
\newblock ``Spin squeezing inequalities for arbitrary spin''.
\newblock \href{https://dx.doi.org/10.1103/PhysRevLett.107.240502}{Phys. Rev.
  Lett. {\bf 107}, 240502}~(2011).

\bibitem{Edmonds1957Angular}
A.~R. Edmonds.
\newblock ``Angular momentum in quantum mechanics''.
\newblock \href{https://dx.doi.org/10.1515/9781400884186}{Princeton University
  Press}. ~(1957).

\bibitem{Toth2004Entanglement}
G\'eza T\'oth.
\newblock ``Entanglement detection in optical lattices of bosonic atoms with
  collective measurements''.
\newblock \href{https://dx.doi.org/10.1103/PhysRevA.69.052327}{Phys. Rev. A
  {\bf 69}, 052327}~(2004).

\bibitem{Toth2007Optimal}
G\'eza T\'oth, Christian Knapp, Otfried G\"uhne, and Hans~J. Briegel.
\newblock ``Optimal spin squeezing inequalities detect bound entanglement in
  spin models''.
\newblock \href{https://dx.doi.org/10.1103/PhysRevLett.99.250405}{Phys. Rev.
  Lett. {\bf 99}, 250405}~(2007).

\bibitem{Toth2010Generation}
G\'eza T\'oth and Morgan~W Mitchell.
\newblock ``Generation of macroscopic singlet states in atomic ensembles''.
\newblock \href{https://dx.doi.org/10.1088/1367-2630/12/5/053007}{New J. Phys.
  {\bf 12}, 053007}~(2010).

\bibitem{Toth2007Detection}
G\'eza {T{\'o}th}.
\newblock ``{Detection of multipartite entanglement in the vicinity of
  symmetric Dicke states}''.
\newblock \href{https://dx.doi.org/10.1364/JOSAB.24.000275}{J. Opt. Soc. Am. B
  {\bf 24}, 275--282}~(2007).

\bibitem{Toth2015Evaluating}
G\'eza T\'oth, Tobias Moroder, and Otfried G\"uhne.
\newblock ``Evaluating convex roof entanglement measures''.
\newblock \href{https://dx.doi.org/10.1103/PhysRevLett.114.160501}{Phys. Rev.
  Lett. {\bf 114}, 160501}~(2015).

\bibitem{Vandenberghe1996Semidefinite}
Lieven Vandenberghe and Stephen Boyd.
\newblock ``Semidefinite programming''.
\newblock \href{https://dx.doi.org/10.1137/1038003}{SIAM Review {\bf 38},
  49--95}~(1996).

\bibitem{Toth2012Multipartite}
G\'eza T\'oth.
\newblock ``Multipartite entanglement and high-precision metrology''.
\newblock \href{https://dx.doi.org/10.1103/PhysRevA.85.022322}{Phys. Rev. A
  {\bf 85}, 022322}~(2012).

\bibitem{Hyllus2012Fisher}
Philipp Hyllus, Wies\l{}aw Laskowski, Roland Krischek, Christian Schwemmer,
  Witlef Wieczorek, Harald Weinfurter, Luca Pezz\'e, and Augusto Smerzi.
\newblock ``Fisher information and multiparticle entanglement''.
\newblock \href{https://dx.doi.org/10.1103/PhysRevA.85.022321}{Phys. Rev. A
  {\bf 85}, 022321}~(2012).

\bibitem{Toth2020Activating}
G\'eza T\'oth, Tam\'as V\'ertesi, Pawe\l{} Horodecki, and Ryszard Horodecki.
\newblock ``Activating hidden metrological usefulness''.
\newblock \href{https://dx.doi.org/10.1103/PhysRevLett.125.020402}{Phys. Rev.
  Lett. {\bf 125}, 020402}~(2020).

\bibitem{Doherty2002Distinguishing}
A.~C. Doherty, Pablo~A. Parrilo, and Federico~M. Spedalieri.
\newblock ``Distinguishing separable and entangled states''.
\newblock \href{https://dx.doi.org/10.1103/PhysRevLett.88.187904}{Phys. Rev.
  Lett. {\bf 88}, 187904}~(2002).

\bibitem{Doherty2004Complete}
Andrew~C. Doherty, Pablo~A. Parrilo, and Federico~M. Spedalieri.
\newblock ``Complete family of separability criteria''.
\newblock \href{https://dx.doi.org/10.1103/PhysRevA.69.022308}{Phys. Rev. A
  {\bf 69}, 022308}~(2004).

\bibitem{Doherty2005Detecting}
Andrew~C. Doherty, Pablo~A. Parrilo, and Federico~M. Spedalieri.
\newblock ``Detecting multipartite entanglement''.
\newblock \href{https://dx.doi.org/10.1103/PhysRevA.71.032333}{Phys. Rev. A
  {\bf 71}, 032333}~(2005).

\bibitem{Ollivier2001Quantum}
Harold Ollivier and Wojciech~H. Zurek.
\newblock ``Quantum discord: A measure of the quantumness of correlations''.
\newblock \href{https://dx.doi.org/10.1103/PhysRevLett.88.017901}{Phys. Rev.
  Lett. {\bf 88}, 017901}~(2001).

\bibitem{Henderson2001Classical}
L.~Henderson and V.~Vedral.
\newblock ``Classical, quantum and total correlations''.
\newblock \href{https://dx.doi.org/10.1088/0305-4470/34/35/315}{J. Phys. A:
  Math. Gen. {\bf 34}, 6899}~(2001).

\bibitem{Bera2017Quantum}
Anindita Bera, Tamoghna Das, Debasis Sadhukhan, Sudipto~Singha Roy, Aditi
  Sen(De), and Ujjwal Sen.
\newblock ``Quantum discord and its allies: a review of recent progress''.
\newblock \href{https://dx.doi.org/10.1088/1361-6633/aa872f}{Rep. Prog. Phys.
  {\bf 81}, 024001}~(2017).

\bibitem{Petz2002Covariance}
D\'enes Petz.
\newblock ``{Covariance and Fisher information in quantum mechanics}''.
\newblock \href{https://dx.doi.org/10.1088/0305-4470/35/4/305}{J. Phys. A:
  Math. Gen. {\bf 35}, 929}~(2002).

\bibitem{Gibilisco2009Quantum}
Paolo Gibilisco, Fumio Hiai, and D{\'e}nes Petz.
\newblock ``{Quantum covariance, quantum Fisher information, and the
  uncertainty relations}''.
\newblock \href{https://dx.doi.org/10.1109/TIT.2008.2008142}{IEEE Trans. Inf.
  Theory {\bf 55}, 439--443}~(2009).

\bibitem{Petz2011Introduction}
D.~Petz and C.~Ghinea.
\newblock ``Introduction to quantum {Fisher} information''.
\newblock \href{https://dx.doi.org/10.1142/9789814338745_0015}{Volume~27, pages
  261--281}.
\newblock World Scientific. ~(2011).

\bibitem{Hansen2008Metric}
Frank Hansen.
\newblock ``Metric adjusted skew information''.
\newblock \href{https://dx.doi.org/10.1073/pnas.0803323105}{Proc. Natl. Acad.
  Sci. U.S.A. {\bf 105}, 9909--9916}~(2008).

\bibitem{Gibilisco2021AUnified}
Paolo Gibilisco, Davide Girolami, and Frank Hansen.
\newblock ``A unified approach to local quantum uncertainty and interferometric
  power by metric adjusted skew information''.
\newblock \href{https://dx.doi.org/10.3390/e23030263}{Entropy {\bf 23},
  263}~(2021).

\bibitem{MATLAB2020}
MATLAB.
\newblock ``9.9.0.1524771(r2020b)''.
\newblock The MathWorks Inc. Natick, Massachusetts~(2020).

\bibitem{MOSEK}
\relax MOSEK~ApS.
\newblock ``{The MOSEK optimization toolbox for MATLAB manual. Version 9.0}''.
\newblock ~(2019).
\newblock
  url:~\href{https://docs.mosek.com/9.0/toolbox/index.html}{docs.mosek.com/9.0/toolbox/index.html}.

\bibitem{Lofberg2004Yalmip}
J.~L{\"{o}}fberg.
\newblock ``{YALMIP : A Toolbox for Modeling and Optimization in MATLAB}''.
\newblock In Proceedings of the CACSD Conference.
\newblock Taipei, Taiwan~(2004).

\bibitem{Toth2008QUBIT4MATLAB}
G{\'e}za {T{\'o}th}.
\newblock ``{QUBIT4MATLAB V3.0: A program package for quantum information
  science and quantum optics for MATLAB}''.
\newblock \href{https://dx.doi.org/10.1016/j.cpc.2008.03.007}{Comput. Phys.
  Commun. {\bf 179}, 430--437}~(2008).

\bibitem{QUBIT4MATLAB_actual_note62_href}
The package QUBIT4MATLAB is available at
  https://www.mathworks.com/matlabcentral/\\fileexchange/8433, and at the
  personal home page https://gtoth.eu/qubit4matlab.html.

\end{thebibliography}

\end{document}